\documentclass[aps,amsfonts,amsmath,prd,preprint,nofootinbib]{revtex4}
\pdfoutput=1
\newcommand{\beq}{\begin{equation}}
\newcommand{\eeq}{\end{equation}}
\usepackage{graphicx}
\usepackage{natbib}

\usepackage{xcolor}

\usepackage{amsmath}
\DeclareMathOperator{\sech}{sech}

\newcommand{\be}{\begin{equation}}
\newcommand{\ee}{\end{equation}}
\newcommand{\bea}{\begin{eqnarray}}
\newcommand{\eea}{\end{eqnarray}}
\newcommand{\bi}{\begin{itemize}}
\newcommand{\ei}{\end{itemize}}
\newcommand{\pa}{\partial}

\begin{document}

\title{The dynamics of Domain Wall Strings}

\author{Jose J. Blanco-Pillado{$^{1,2,3}$}\footnote{josejuan.blanco@ehu.eus}, Daniel Jim\'enez-Aguilar{$^{1,2}$}\footnote{daniel.jimenez@ehu.eus}, Jose M. Queiruga{$^{4,5}$}\footnote{xose.queiruga@usal.es} and Jon Urrestilla{$^{1,2}$} \footnote{jon.urrestilla@ehu.eus}}

\affiliation{$^1$ Department of Physics, University of Basque Country, UPV/EHU, 48080, Bilbao, Spain \\
$^2$ EHU Quantum Center, University of Basque Country, UPV/EHU\\
$^3$ IKERBASQUE, Basque Foundation for Science, 48011, Bilbao, Spain \\
$^4$ Department of Applied Mathematics,
University of Salamanca, 37008, Salamanca, Spain \\
$^5$
Institute of Fundamental Physics and Mathematics,
University of Salamanca, 37008 Salamanca, Spain.
}

\begin{abstract}

We study the dynamics of domain wall solitons in $(2+1)d$ field theories. 
These objects are extended along one of the spatial directions, so they
also behave as strings; hence the
name of domain wall strings. We show analytically and numerically that 
the amount of radiation from the propagation of wiggles on these objects 
is negligible except for regions of high curvature. Therefore, at low
curvatures, the domain wall strings behave exactly as the Nambu-Goto action
predicts. We show this explicitly with the use of several different 
numerical experiments of the evolution of these objects in a lattice. We 
then explore their dynamics in the presence of internal mode
excitations. We do this again by performing field theory simulations and identify an effective action 
that captures the relevant interactions between the different degrees of
freedom living on the string. We uncover a new parametric resonance instability that
transfers energy from the internal mode to the position of the
domain wall. We show that this instability accelerates the radiation
of the internal mode energy. We also explore the possibility of
exciting the internal mode of the soliton with the collision of wiggles
on the domain wall. Our numerical experiments indicate that this
does not happen unless the wiggles have already a wavelength 
of the order of the string thickness. Finally, we comment on the possible
relevance of our findings to cosmological networks of defects. We argue 
that our results cast some doubts on the significance
of the internal modes in cosmological applications beyond a brief
transient period right after their formation. This, however, should be
further investigated using cosmological simulations of our model.

\end{abstract}

\maketitle

\section{Introduction}

Solitons emerge as non-perturbative solutions of field theory models in
many branches of physics from condensed matter to
particle physics or cosmology. In many
cases, these solitons are created at a symmetry breaking phase transition.
The energy scale associated with this transition is therefore the characteristic
scale of the problem when studying these objects. Their mass as well as their
size is controlled by this scale. This also suggests that the typical excitations of
these solutions should relax to the lowest energy configuration on a time 
of the order of the characteristic scale of the soliton, the light crossing time
of the object. All this seems to indicate that if we are only interested in studying 
these objects at low energies or for long periods of time, we can forget about 
these excitations. 

However, the study of field theory perturbations around these stable 
solutions reveals that in many cases there are long lived excitations
spatially localized around the soliton \cite{Arodz:1991ws, Goodband:1995rt,Alonso-Izquierdo:2015tta, Blanco-Pillado:2020smt, Blanco-Pillado:2021jad}. 
This extra energy gives rise to a
new kind of object, the excited soliton, whose macroscopic behaviour 
could be altered by the presence of these excitations. Furthermore,
the existence of these localized modes points to an interesting coupling
of the soliton to the radiation modes.  This coupling enables the soliton to
slowly relax to the lowest energy configuration but also indicates the possibility
of interactions with the environment.  All these processes could in principle
become important in the study of the dynamics of these objects.

On the other hand, depending on the particular model and the dimensionality of 
spacetime, these solitons could appear as extended objects such as
strings or domain walls in $(3+1)$ dimensions. The presence of these longitudinal directions
allows for possible excitations of the soliton to propagate along them. For a particular
model there could be more than one type of these excitations. For example, in some 
models one could have waves of the localized modes described earlier travelling
along the objects. In addition, the breaking of translational invariance by the
object guarantees the appearance of Goldstone modes living on the worldvolume
of these solitons. In general, one would have perturbations of all these
modes moving up and down the soliton, interacting among themselves
and possibly emitting or absorbing radiation. It is therefore clear that in order to 
understand the dynamics of these objects one would need to find the conditions
when these excitations become important and obtain the effective 
theory that governs these modes and their interactions.

A traditional approach to this problem has used the fact that there is
a separation of scales between the different sets of perturbations
present in these objects. The localized excitations on the core of the solitons
have typically masses comparable with the characteristic scale of the field
theory that gives rise to them. However, the perturbations that represent
wiggles on the solitons are, by definition, massless from the point
of view of the effective theory living on the worldvolume. Therefore, a reasonable 
strategy to find an approximate description of the soliton's dynamics would be to identify 
the action that captures the interactions of the low energy degrees of freedom, in 
this case, the massless modes. This assumes that the massive modes 
are not easily excited by the subsequent evolution since the energy
required to excite them is not present at late times.

In the case of strings, this approach leads to the use of the so-called Nambu-Goto (NG)
action \cite{osti_4118139,Goto:1971ce,NIELSEN197345}. This action is basically the generalization to the case of a string of the 
relativistic particle action and it is described by the area of the worldsheet of 
the string \footnote{Note that this is also an approximation on the more general 
action that would include corrections due to the thickness of the string.}. Taking 
this viewpoint, one can develop large scale simulations that track the motion 
of a network of strings. This is particularly interesting in the context of cosmology, 
where these relativistic strings have been predicted in many extensions of the 
Standard Model \cite{Kibble:1976sj,Jeannerot:2003qv}.

Alternatively, one can simulate the dynamics of these objects evolving the full field theory 
in a lattice. This method follows all the degrees of freedom and in principle one 
includes all the interactions between them. However, the obvious price to pay 
in this case is the limited dynamic range that one can have in these simulations.
Resolving all the field theory structure in the solitons at scales smaller than their
thickness means that one can not simulate as many strings as in the Nambu-Goto
case.

It has always been assumed that both methods should be reconciled
in some regime demonstrating their complementarity to study the dynamics of these
strings. However, some of the results from field theory simulations of local cosmic
string networks deviate from their NG counterparts. In particular, these lattice
simulations seem to indicate that loops (closed strings) produced during their cosmological
evolution quickly collapse and disappear. This is in contrast with the results in NG dynamics,
where one is expected to have some long lived oscillating loops.
This apparent mismatch between both types of simulations is 
even more puzzling taking into account that simulations of individual strings in 
field theory do indeed follow the NG dynamics \cite{Olum:1998ag,Olum:1999sg}.  Understanding this
puzzling issue has become one of the central problems in the cosmic string community, specially 
since it is of paramount importance in the calculation of the expected gravitational 
wave signatures of these models \cite{Blanco-Pillado:2017oxo,Auclair:2019wcv}.

Recently, it has been suggested that this different behaviour of string loops
could be due to the presence of excited strings in the field theory simulations \cite{Hindmarsh:2017qff,Hindmarsh:2021mnl}. Moreover,
it was also conjectured that these excitations could be in the form of localized 
bound modes on the strings that we mentioned earlier. This argues for a possible extension
of the effective theory of the string motion by including these extra massive modes.
Exploring this possibility in detail in a realistic $(3+1)$ dimensional simulation is quite 
difficult, so here we will investigate this issue in a simpler setting, in $(2+1)$d.
Even though the problem that we study in the present paper is quite 
different from the actual situation for strings, it shares many of the issues
that are at play in that case. We will study domain wall solitons that in 
$(2+1)$ dimensions behave as string-like objects, thus the name {\it domain wall strings} in our title. 
Their effective action within the thin wall approximation is 
identical to the Nambu-Goto action for strings. Furthermore,
we will demonstrate that the walls in these simple scalar field models have 
bound state excitations stuck to their core that decay slowly into radiation.
Therefore, the lower dimensional problem that we will study here has all the 
ingredients that have been presented in the hypothetical resolution of the local cosmic string
puzzle. Taking all this into account, we hope that some of our results and conclusions can be extrapolated to the
actual problem with local strings.

The organization of the paper is the following. In Section II we will describe
the field theory model that gives rise to the domain wall solutions that we are 
interested in studying and discuss the different types of fluctuations
that one can obtain around these solutions. In section III we will discuss the
different processes that lead to energy loss by the domain wall dynamics. In Section IV we discuss the 
dynamics of these domain wall strings in the absence of any internal excitation,
paying particular attention to the region of validity of the thin wall approximation,
the Nambu-Goto equations of motion. In Section V we include the presence
of massive excitations on the effective action and discuss their possible
dynamical implications. In Section VI we explore the non-linear interactions
between the excitations on the string and their Goldstone modes. We 
demonstrate the existence of resonance effects between them and the
transfer of energy between these modes as well as their radiation. In Section
VII we study the possible generation of the bound state mode from the
interaction of Goldstone modes. Finally, we summarize our results in the conclusions and
give a brief description of the relevance of these effects for cosmological networks.\\

Some examples of the simulations we have performed in this paper can be found at
http://tp.lc.ehu.es/earlyuniverse/dynamics-of-domain-wall-strings.

\section{Field theory model for domain walls and their excitations}

The field theory model that we will investigate in this paper is described by the following
$(2+1)$ dimensional action for a scalar field $\phi(x,y,t)$:
\beq\label{Act:DW}
S = \int{d^3x \left[\frac{1}{2} \partial_{\mu} \phi \partial^{\mu} \phi - \frac{\lambda}{4} \left(\phi^2 - \eta^2\right)^2\right]}~,
\eeq
whose equations of motion are therefore given by
\beq
\partial_{\mu} \partial^{\mu} \phi + \lambda \phi \left(\phi^2 - \eta^2\right) = 0 ~.
\eeq

Looking at this equation, one can immediately identify the existence in this theory of two degenerate
vacua at $\phi = \pm \eta$. Small fluctuations about each of these vacua give
rise to the perturbative excitations of the theory, whose masses are given by
$m^2 = 2 \lambda \eta^2$. In other words, the theory we will consider has a
gap since there are no massless propagating modes in the $2+1$ dimensional vacua of this theory.
This is important for our considerations since, as will discuss later on, this
will be the only available channel for decay of any excitation present on the solitons.
This, in turn, will determine the different time scales associated with those
excitations.

\subsection{The domain wall string solution}

It is well known that this theory possesses, as part of its spectrum,
non-perturbative solitonic solutions that interpolate between the two vacua. These solitons can easily
be obtained as static solutions of the equations of motion of the theory and
take the following functional form:
\beq
\label{kinksolution}
\phi_K(x,y,t) = \eta \tanh\left(\frac{m}{2} x\right)~.
\eeq

The properties of these solutions are well known and have been
studied in many papers before (see for example the review in \cite{Vachaspati:2006zz}). Here 
we will briefly summarize the most important results relevant for the rest of our discussion.

The first important result is about stability. One can show that these configurations
are stable once one imposes the asymptotic boundary conditions for $x\rightarrow \pm \infty$.
This property is inherited from the analogue solution of the $1+1$ theory, the kink solution, and represents the
prototypical example of topological stability \footnote{This does not mean that all configurations constructed with
these solutions are stable. In $2+1$ dimensions one can easily build configurations, like a circular loop of the 
kink soliton, that will be unstable to collapse. We will discuss both types of solutions in our paper.}.

Furthermore, the configuration in Eq. (\ref{kinksolution}) does not depend on one of the coordinates, in our case the $y$ direction.
This means that one should think about this soliton as an extended object along that direction.
It is easy to check that most of the energy density on these field solutions is concentrated
within a narrow area of thickness $\delta  \sim m^{-1}$
 around the $\phi=0$ region (the core).
Taking this into account, we can consider this object as a line-like defect
in our lower dimensional spacetime in ($2+1$)d. This means that even
though these field configurations are domain walls that separate two
different vacua in our model they will dynamically behave as strings. That is why
we named them {\it domain wall strings}.

Finally, given the solution we can compute the equation of state associated to these
string-like objects. Given the fact that these field configurations are independent of the
$y$ direction as well as time, we can see that these domain wall strings will have
a relativistic string equation of state where the energy per unit length is equal to the
pressure along the $y$ direction, the tension of the string. One can compute this
energy per unit length analytically in our model and find that \footnote{Note that in $2+1$ dimensions
the parameter $\lambda$ has units of energy.}
\beq
\mu =  \left( \frac{m}{3\lambda}\right) m^2~.
\eeq

\subsection{Spectrum of excitations}
\label{spectrum}

We can now study the spectrum of small perturbations around the static solution
given by Eq. (\ref{kinksolution}). In order to do this, we will start by considering the
following ansatz for these perturbations \footnote{Here we closely follow
the discussion given in \cite{Blanco-Pillado:2020smt}, where the study of this spectrum was done in a lower dimensional
setup.}:
\beq
\phi(x,y,t) = \phi_K(x) + f_n (x) e^{i (k_n y - \omega_n t)}~.
\eeq
Inserting this expression in the equations of motion for the scalar field we obtain,
at the linear level, the following equations:
\beq
\label{Schrodinger-eq}
-f''_n (x) + U(x) f_n(x) = \Omega^2_n f_n(x)\,,
\eeq
where
\beq
U(x) = \lambda \left(3 \phi_K^2(x) - \eta^2 \right) 
\eeq 
is the effective potential for this Schr\"odinger problem and the associated eigenvalues are 
described in terms of the parameters in the ansatz as
\beq
\Omega_n^2 = \omega_n^2 - k_n^2 ~.
\eeq

It turns out that the eigenvalue problem given by Eq. (\ref{Schrodinger-eq}) can be exactly solved \cite{Rajaraman:1982is}. Its spectrum
is given in terms of two bound states whose properly normalized eigenfunctions are
\beq
f_0(x) = \sqrt{\frac{3m}8} \sech^2\left(\frac{mx}2\right) ~~~~\text{with} ~~~~~ \Omega_0 =  0
\label{eq:zm}
\eeq
and
\beq
f_1(x) = \sqrt{\frac{3m}4} \sinh\left(\frac{mx}2\right) \sech^2\left(\frac{mx}2\right) ~~~~\text{with} ~~~~~ \Omega_1=  \frac{\sqrt{3}}{2}m\,,
\label{eq:sm}
\eeq
and a continuum of states starting at $\Omega_n = m$.

Each of these modes has a simple physical interpretation which follows from their
lower dimensional analogues \cite{Blanco-Pillado:2020smt}. The modes characterised
by $\Omega_0 =  0$ represent the local rigid displacements of the domain wall profile
that propagate at the speed of light along the longitudinal directions of the object,
the homogeneous one being just a rigid displacement of the whole object.

The modes associated with the other bound state, with $\Omega_1=  \frac{\sqrt{3}}{2}m$,
represent travelling excitations of the so-called {\it internal mode} or {\it shape mode}.  One can think of
it as a perturbation in the width of the soliton that oscillates in time with frequency
$\Omega_1$. 

\begin{figure}[h!]
\includegraphics[width=8cm]{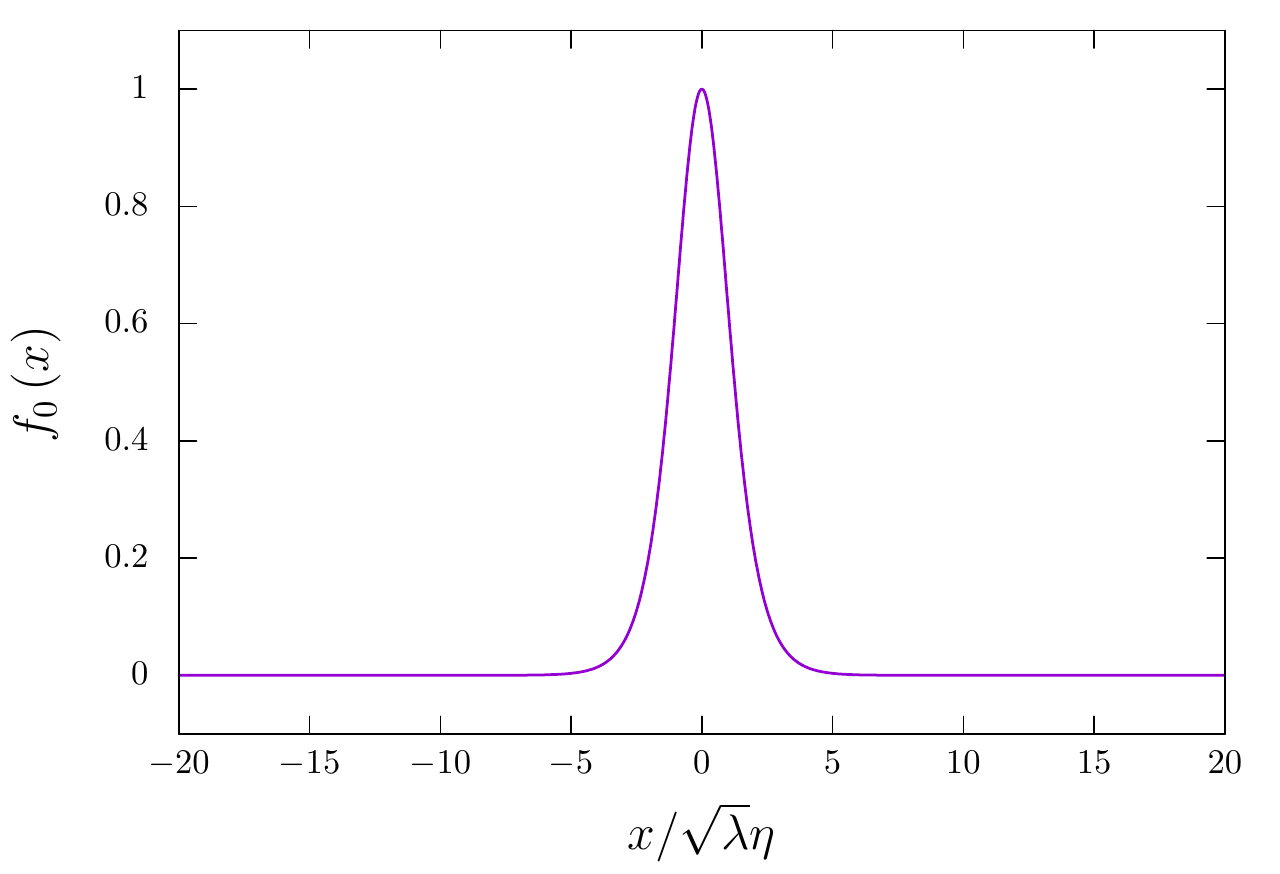}
\includegraphics[width=8cm]{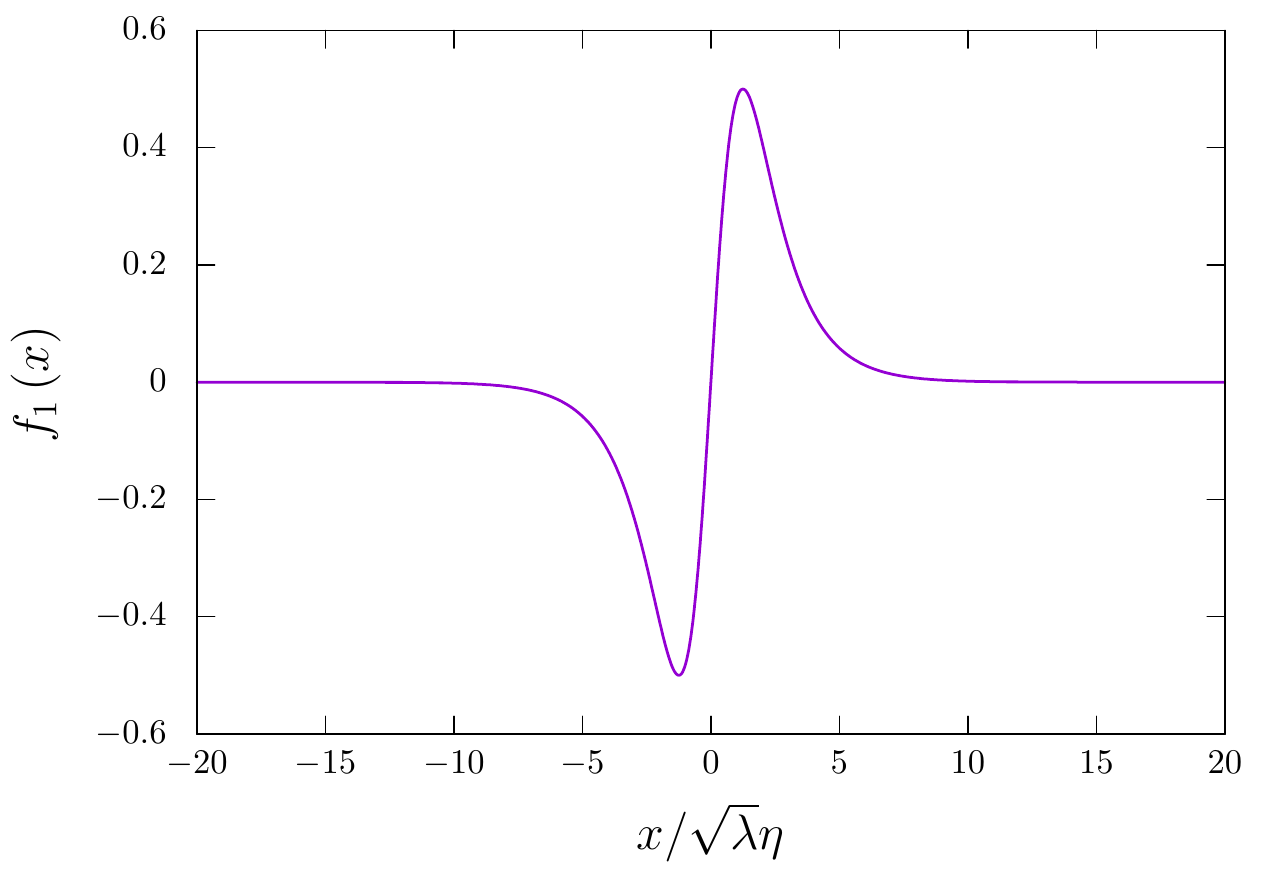}
\caption{Zero mode (left panel) and shape mode (right panel) excitations.} 
\label{fig:zero mode and shape mode}
\end{figure}

The profiles of the zero mode and the shape mode, given respectively in Eqs. (\ref{eq:zm}) and (\ref{eq:sm}), are shown in Fig. \ref{fig:zero mode and shape mode}. We note that
both these modes have a spatial extent comparable to the size of the soliton itself, namely, $\delta 
\sim m^{-1}$.

Finally, there is also the continuum of scattering modes starting at frequency $m$ that represent all the 
excitations of the field outside of the soliton; in the vacuum. A generic field configuration
would be described as a combination of all these modes with their corresponding 
amplitude. In the following, we will try to understand how to describe the dynamics of
the domain wall string in the presence of some of these perturbations and their possible interactions.

\section{Radiation from domain wall excitations}

As we described in the introduction, one of our goals is to identify the appropriate
dynamics that characterises the domain wall string motion and its coupling to the internal
modes. This task would be greatly simplified if we could write an effective action
for all the modes populating the string in a $2d$ worldsheet theory. However, this type
of action disregards the possibility of energy leaking into the bulk, so in order
to be able to quantify the regime of validity of this approach, one should
understand the presence of radiative processes first.

\subsection{Radiation from zero mode excitations}

As we alluded earlier, the zero mode excitations represent the Goldstone
modes associated with the wiggles of the domain wall string. A very interesting
result due to Vachaspati and Vachaspati \cite{Vachaspati:1990sk} shows that one can obtain the
fully non-linear description of these excitations at the field theory level. In particular,
assuming a domain wall extended along the $y$ direction, one can show that
the configuration of the form
\beq
\phi(x,y,t) = \phi_K \left(x - \psi(y \pm t)\right)
\eeq
is an exact solution of the domain wall field theory with an arbitrary transverse excitation of the form
$\psi(y \pm t)$. This wiggle moves at the speed of light in one direction along the domain wall 
and does not change its shape. The energy is conserved, so no radiation
is being emitted from these waves. However, the most general configuration 
of the zero mode excitations will have in general fluctuations propagating in both
directions. In this case, we expect some radiation to be emitted from the domain wall.

In order to study this radiation in a quantitative manner, we will start by investigating the 
energy loss from a particular set of perturbations described by 
standing waves where the center of the domain wall is parametrized by
\beq
\psi_0(y,t) = \hat{D} \cos( \omega_{0}y)\cos( \omega_{0}t)~.
\eeq
A configuration of this type is not an exact solution. However,
for small enough amplitude, we expect these waves to behave like
a free field on the worldsheet, as the linear theory predicts. Their oscillation will
then act as a source for radiation at a quadratic order in their amplitude. It is this
radiation that we want to study. 

We compute in the Appendix \ref{an_zero} the approximated 
power emitted per unit length of the domain wall string from these configurations 
as a function of the frequency $\omega_{0}$. In order to perform this
calculation we have assumed that the scalar field profile during these
oscillations is well approximated by the ansatz
\beq
\label{FT-for-NG}
\phi(x,y,t) = \phi_K\left( \frac{x- \psi_0(y,t)}{\sqrt{1- \partial_a \psi_0 \partial^a \psi_0}}\right)\,,
\eeq
where $a=(y,t)$ are the coordinates on the worldsheet of the domain wall string.
We demonstrate in the Appendix \ref{appendix-ansatz for NG dynamics} that this ansatz is a good
approximation for the full field theory equations as long as the
second derivatives of the function $\psi(y,t)$ are small \footnote{ As we note in the Appendix \ref{appendix-ansatz for NG dynamics}, 
this ansatz for the field profile is the correct one for a string moving according to the
Nambu-Goto action as we expect for these types of waves. We will have much more to say 
about this expectation in the coming sections on the validity
of the Nambu-Goto action later on in the paper. We could have also taken the ansatz
\beq
\phi(x,y,t) = \phi_K(x) + \psi_0(y,t)~\times f_0(x)~,
\eeq
which is the linear approximation of the one we used. The results in this case are qualitatively 
similar to the ones presented here. }.

\begin{figure}[h!]
\includegraphics[width=14cm]{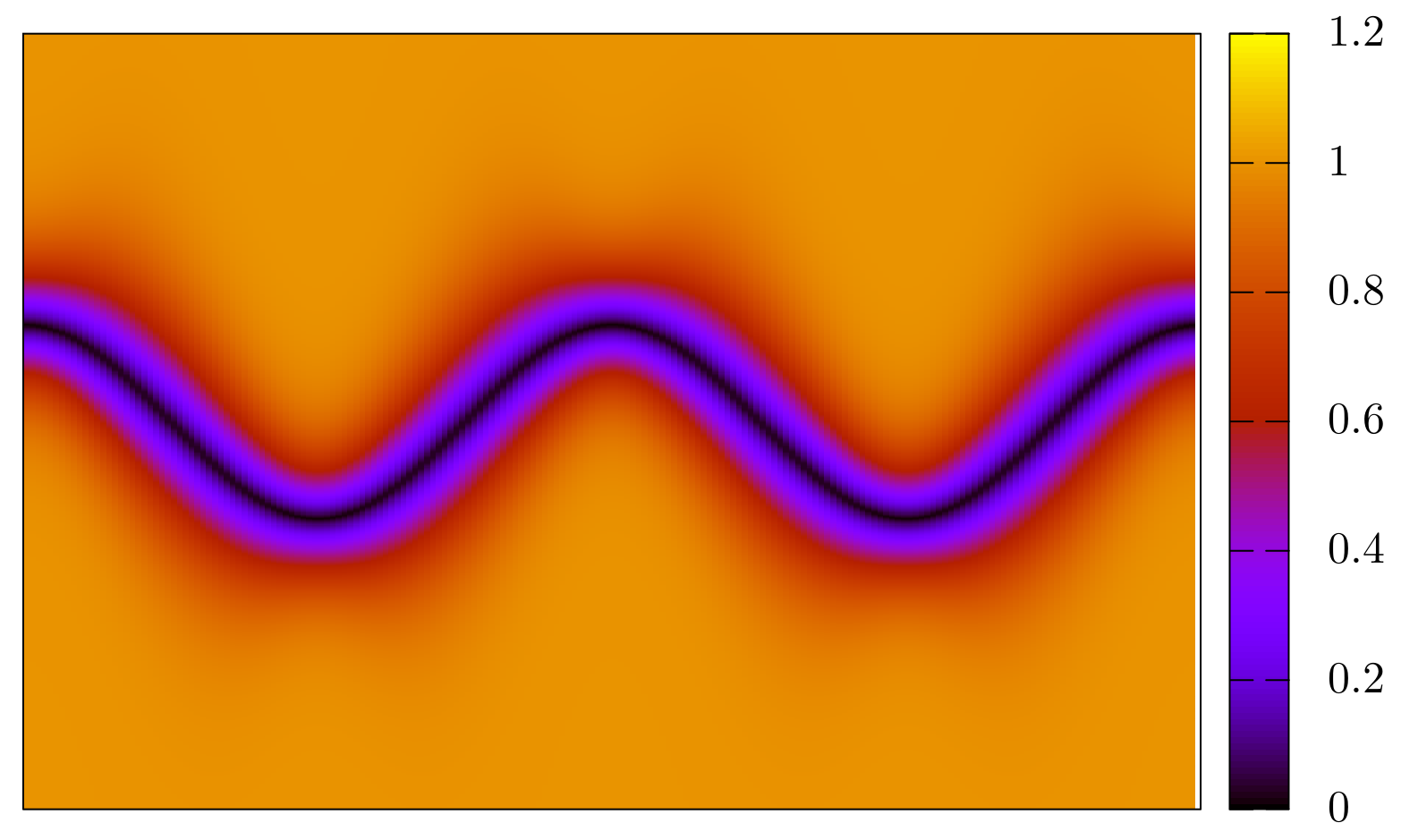}
\caption{Initial configuration (\ref{eq:initial conditions 3a}). The color palette indicates the value of $|\phi|/\eta$.} 
\label{fig:figure initial condition 3a}
\end{figure}

On the other hand, we have also explored this process in a field theory lattice simulation. 
Taking the ansatz given above, we can initialize the numerical simulations
for a sinusoid excitation of a particular wavelength and amplitude:
\begin{equation}
\phi\left(x,y,0\right)=\phi_{K}\left(\frac{x-\hat{D}\cos\left(\omega_{0}y\right)}{\sqrt{1+\hat{D}^{2}\omega_{0}^{2}\sin^{2}\left(\omega_{0}y\right)}}\right),\,\,\,\,\,\,\,\,\,\,\dot{\phi}\left(x,y,0\right)=0\,.
\label{eq:initial conditions 3a}
\end{equation}
This initial state has the form shown in Fig. \ref{fig:figure initial condition 3a}.

Evolving this configuration in a box with absorbing boundary conditions
on the boundaries transverse to the string, we can obtain the power
emitted at each moment in time directly from the simulation \footnote{See the
details of the field theory techniques that we use throughout this paper in Appendix \ref{appendix-numerics}.}. We can
actually do this in two different ways: either
by looking at the decrease of the total energy in the box or by
integrating the energy flux through a surface near the boundary.
Both strategies yield consistent results.

\begin{figure}[h!]
\includegraphics[width=14cm]{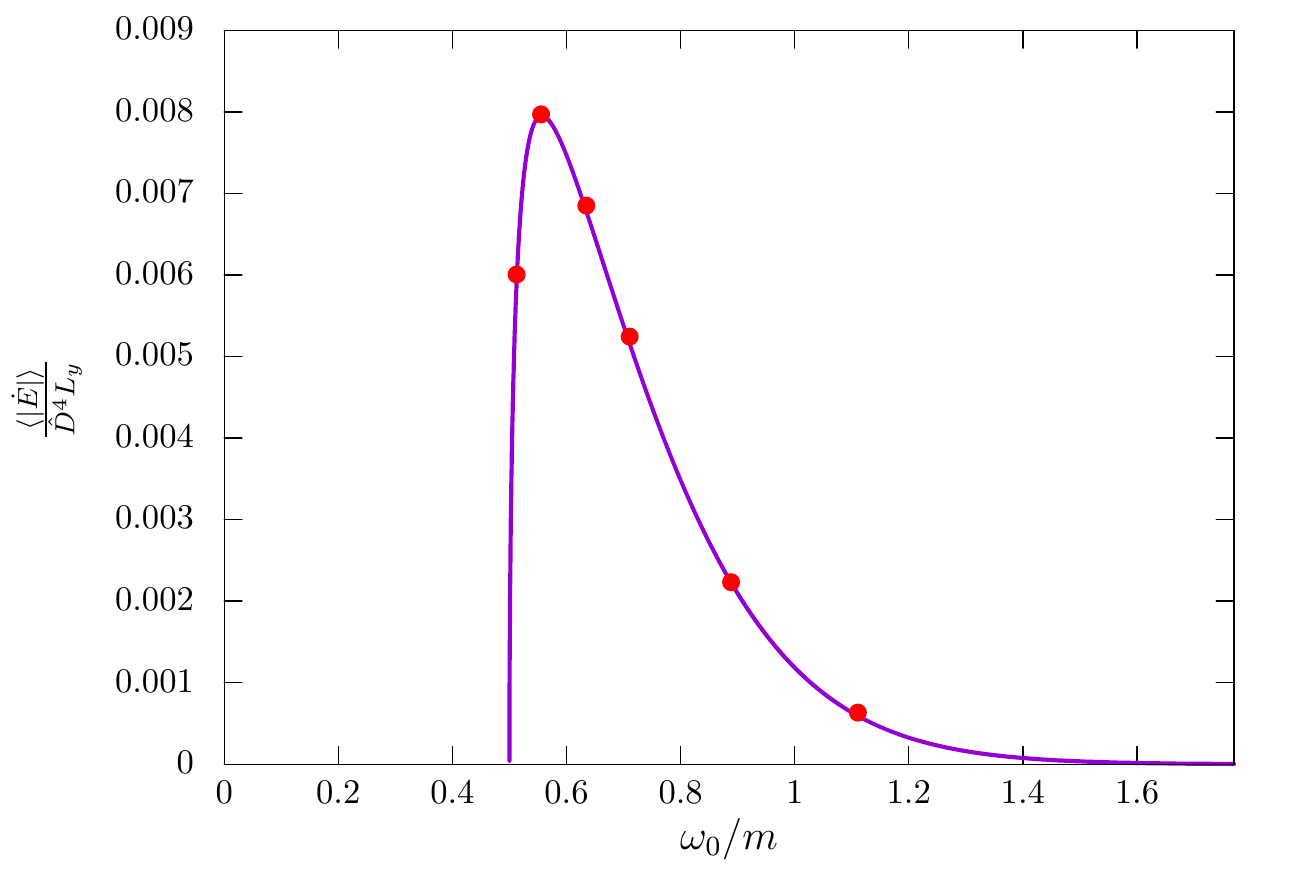}
\caption{Radiated power per unit length as a function of the frequency of the zero mode. 
The quantity on the $y$ axis is made dimensionless by dividing the power by $\lambda^{3}\eta^{8}$. The solid curve corresponds to the analytical estimate
(\ref{eq:average power per unit length}), and the red points represent the power read directly
 from the numerical simulations.} 
\label{fig:power zero mode}
\end{figure}

We show in Fig. \ref{fig:power zero mode} the comparison of this power computed using the
analytic expression and the numerical simulations. The results
show a very good agreement between these two approaches.

These results indicate several important facts for us. First and
foremost, it is clear that for a frequency below $\omega_{*} = \frac{m}{2}$,
the radiation is highly suppressed. In fact, it is easy to see in the
analytic calculation that the quadratic order coupling shuts off
at frequencies below this threshold value. This can be easily understood
by recognizing that below this frequency the quadratic source would
oscillate with a frequency below the lowest propagating mode in
the vacuum. This indicates that for low amplitudes and for 
frequencies below $\omega_{*}$ one should not expect any
appreciable radiation \footnote{This is, of course, not strictly true
since there should be some higher order radiation. However, at
low enough frequencies, this should be negligible since the
radiation would be quartic on the amplitude.}.

Increasing the frequency of oscillation beyond this threshold,
the string starts radiating slowly and its power grows with frequency. 
However, this will also shut down for high enough frequency. This 
effect is basically due to the finite size of the source, which in our
model is also parametrized by the mass of the particle being 
radiated ($m$). This peculiar behaviour is due to the fact that
both scales, the mass of the particle radiated and the inverse
of the size of the source, are of the same order.

We can also take this type of ansatz and extend it beyond
its range of expected validity to study waves of high amplitude and curvature.
In this case, a new source of radiation appears that has to do
with the regions of high curvature. The standing wave solutions
we are discussing will have their highest curvature at their
maximum extension. We have observed that these configurations 
seem to radiate much more intensively
in the case where the associated radius of curvature is of the
order of the domain wall thickness. This is not surprising 
since in this case we expect the different segments of the
string to start interfering with one another since they
are forced to be closer than their own thickness. This is
reminiscent to what happens at the high curvature regions
in the case of local cosmic strings \cite{Olum:1998ag,Olum:1999sg}.

\begin{figure}[h!]
\includegraphics[width=14cm]{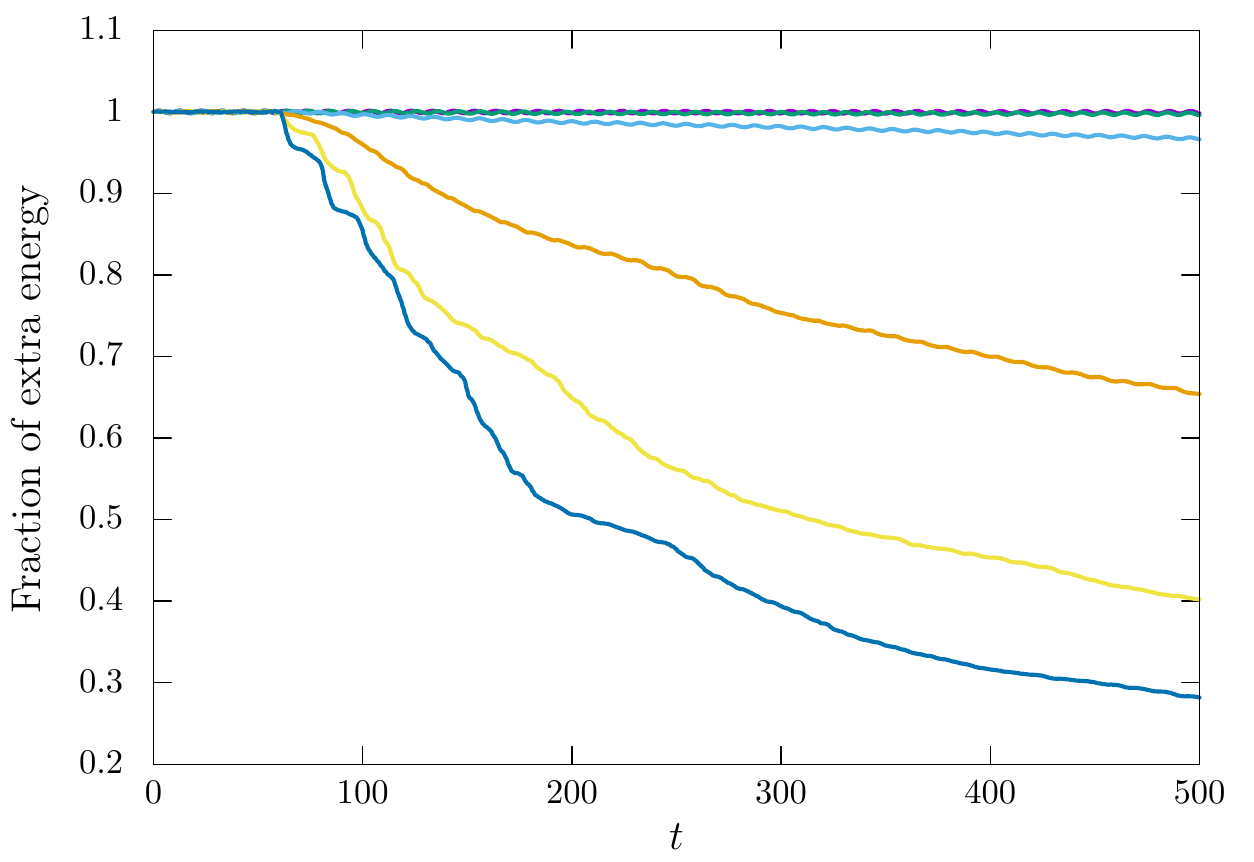}
\caption{Fraction of extra energy as a function of time, given in units 
of $\left(\sqrt{\lambda}\eta\right)^{-1}$, for standing waves with increasing 
amplitudes. ``Extra energy'' means total energy inside the box minus energy 
of the straight domain wall. From top to bottom, the amplitudes of the standing 
waves are $\hat D=0.5, 1, 2, 4, 6,$ and $8$, with corresponding curvatures (i.e., ratio 
of string thickness to radius of curvature) $\kappa=0.07, 0.14, 0.28, 0.56, 0.84,$ 
and $1.12$. In all cases, the frequency of the standing wave is $\omega_{0}<m/2$.} 
\label{fig:nonperturbative radiation}
\end{figure}

In Fig. \ref{fig:nonperturbative radiation} we illustrate how the standing 
wave loses energy in the form of non-perturbative radiation. We choose 
an angular frequency below $m/2$ so that the zero mode does not couple 
quadratically to the scattering states, and look at the power emitted by the 
standing wave as we gradually increase its amplitude. We see that for
low amplitudes the radiation is negligible and it is enhanced for large amplitudes.

We can also describe the different configurations by specifying their ratio between 
the domain wall string thickness and the radius of curvature at the crests of 
the waves, namely the quantity
\beq
\kappa = \frac{\delta}{(\hat D \omega_0^2)^{-1}}\,.
\label{def-curvature}
\eeq
 We note that the radiated power is only significant for large amplitudes
 where $\kappa \ge 0.5$. This seems to be in agreement with the
 expectations that this energy loss mechanism has a non-perturbative 
 origin.
 
\subsection{Radiation from internal mode excitations}

As we described before, a homogeneous internal excitation on the domain wall (one that is independent
of the $y$ coordinate) would lead to an oscillation of the soliton thickness\footnote{One can see this by looking at the profile of the internal mode given in Fig. \ref{fig:zero mode and shape mode}.} with a frequency such that $2\Omega_1 > m$. This
tells us that non-linear effects will lead to radiation since these excitations can act as a
source of modes that are allowed to propagate in the bulk.

 This effect was extensively studied in numerical simulations
in $1+1$ dimensions in \cite{Blanco-Pillado:2020smt} for kinks
excited with the internal or shape mode. Here we will just outline the most important
findings of that paper.

Assuming that one starts with a homogeneous excitation of the bound state
of the form
\beq
\phi(x,y,t) = \phi_K(x) + \hat A(t) \cos(\Omega_1 t) \times f_1(x)~,
\eeq
one can show \cite{Manton:1996ex,Blanco-Pillado:2020smt} that its amplitude
will follow a time dependent expression of the form

\beq
\hat A(t)^{-2} = \hat A(0)^{-2} \left( 1+  \frac {t}{\tau}\right)\,,
\eeq
where 
\beq
\tau =  \left(\frac{\Gamma }{ \hat A(0)^{2} } \right) m^{-1}\,.
 \eeq
$\Gamma$ is a numerical coefficient that can be analytically computed
from the kink profile and is of ${\cal O} (100)$. This expression demonstrates that the shape mode 
amplitude decreases slowly compared to the time scale associated with the thickness of the 
soliton. This justifies the interest on the amplitude of these modes in field theory
simulations, which can be running for a short time compared to this decay time.

The physical reason for this slow decay of the shape mode can be traced back
to its non-linear origin. At the linear level the shape mode is stable, but
with the non-linear interactions included, one can see the presence of
radiation that slowly depletes the energy of this mode. This was shown
to be the case explicitly in simulations performed in \cite{Blanco-Pillado:2020smt}.

We can also look at the evolution of more complicated configurations where
the shape mode is excited in the form of a standing wave with some finite 
wavelength along the $y$ direction. The analytic estimates presented in Appendix \ref{an_shape} show that for frequencies not
too large compared to the mass of the propagating particle in the bulk, the associated 
decay time scale for these modes is comparable to the homogeneous one \footnote{We can also think about travelling waves of this mode moving in one
direction. However, these configurations can be
brought to look like the homogeneous one by a boost along the longitudinal
direction of the domain wall. Therefore, we could estimate the new time scale
associated to their decay by performing this boost on the homogeneous mode.}.

Finally, similarly to what happens in the case of the zero modes, we should 
always keep in mind that these results are accurate when one
considers small amplitude perturbations. Going beyond this will introduce
errors that could lead to drastically different evolutions of the field. This, for
example, limits the amplitude of the homogenous shape mode to be below unity. At
high amplitudes, the soliton will not be able to sustain the excitation and
can lead to pair creation of solitons \cite{Blanco-Pillado:2020smt}\footnote{You can
also  look at the simulations in http://tp.lc.ehu.es/earlyuniverse/kink-simulations }. \\

All these considerations suggest that we could try to write an effective
action for the excited domain walls for small curvatures and time scales $t<\tau$.
In this regime, the amount of radiation from the domain wall should be sub-dominant
and one could try to find the relevant couplings between the different
modes on the worldsheet. That action would allow us to explore the 
mechanical backreaction on the domain walls due to the presence of the 
shape mode. This could lead to different dynamics of the domain walls
and perhaps explain the different behaviour of strings in field theory simulations. 
However, before we do that, let's start by identifying in the next section the correct action 
in the absence of any internal excitation. After that, we will 
add new terms to the action that capture the new dynamics and couplings 
between the zero and internal modes.

\section{The dynamics of bare domain wall strings}
\label{NG-section}

Let's consider for a moment the low energy dynamics of our domain walls
in $(2+1)d$. At these low energies the internal mode will not get excited,
so one can assume that the only dynamical degree of freedom is the position
of the wall. Furthermore, we will also assume that there are no high curvature
regions excited on the domain wall, so we can neglect
the effects of radiation.

The evolution of the wall can then be described by a 
$2d$ worldsheet specified by the 3-vector $X^{\mu}(\xi^0,\xi^1)$, where
$\xi^a$ are the internal coordinates of the worldsheet with $(a = 0,1)$. The simplest
possible action describing the dynamics of these objects can
be found by generalizing the relativistic point particle action to
a line-like object. Following this idea, we conclude that the action of a relativistic 
string-like object should be identified with the area of its worldsheet, namely,
\beq
S_{NG} = -\mu \int{d^2 \xi \sqrt{-\gamma}}\,,
\eeq
where $\gamma$ denotes the determinant of the induced metric on the worldsheet and
$\mu$ is the energy per unit length of the object. This is the so-called Nambu-Goto
action  \cite{osti_4118139,Goto:1971ce,NIELSEN197345}.

It is clear that this action is invariant under reparametrizations of the 
worldsheet coordinates. This is good since it means that the physics 
does not depend on the way we choose our coordinate system on the domain wall
string worldsheet. This also means that there are many different gauges that one
can adopt to describe the dynamics of the wall. The equations of motion
in any gauge are given by
\beq
\label{general-eqs}
\partial_a\left(\sqrt{-\gamma} \gamma^{ab} X^{\mu}_{,b}\right) = 0~,
\eeq
where the induced metric is given by
\beq
\gamma_{ab} = \eta_{\mu \nu} \partial_a X^{\mu} \partial_b X^{\nu}~.
\eeq

Our goal in this section of the paper is to compare the evolution
of the field theory solitons in $2+1$ dimensions we presented earlier with the solutions
predicted by these equations. In order to do that, we will design several
numerical experiments that will allow us to understand
the physical conditions that should be met for these two descriptions
of the same physics to agree.

The equations of motion for the Nambu-Goto string wall written above 
look quite complicated. In particular, expanding them in terms of our 
variables, the functions $X^{\mu}$, we would obtain a scary-looking
set of non-linear differential equations. We could, of course, solve them 
numerically as well. However,  the situation is greatly simplified by a judicious 
choice of gauge.

Let us start our discussion with the usual conformal gauge\footnote{For more details,
see \cite{Vilenkin:2000jqa}.}. In this
case, one identifies the timelike coordinate with the zero component
of the string position, $X^0 = t$, and uses the gauge freedom to impose the following conditions
on the 2-vector ${\bf X}$ that denotes the position of the string:
\bea
{\bf \dot X} \cdot {\bf X'} &=&0 ~,\\
{\bf \dot X}^2 + {\bf X'}^2 &=& 1\,.
\eea
Here, and in the rest of the paper, the derivatives with respect to time are denoted by a dot and the primes describe the derivatives with respect to the spacelike worldsheet coordinate,
which is normally represented by $\sigma$.

It is worth explaning the physical meaning of these conditions. The first one
tells us that the velocity of the string will always point in the direction
perpendicular to its local tangent vector, while the second one fixes
$\sigma$ to parametrize the energy along the domain wall string. The interesting thing
about this gauge is that the equations of motion become 
\begin{equation}
{\bf \ddot X} - {\bf X''}=0\,,
\end{equation}
which is simply the wave equation for the  2-vector, ${\bf X}(\sigma, t)$. The solution
of these equations is well known and can be written in terms of right-movers
and left-movers along the wall as
\begin{equation}
\label{NG-sol}
{\bf X}(t,\sigma) = \frac{1}{2}\left( {\bf a} (\sigma-t) +{\bf b} (\sigma + t)  \right)~,
\end{equation}
where in order to fulfill the gauge conditions the functions ${\bf a}'$ and ${\bf b}'$
must satify
 \beq
 \label{constraints}
 | {\bf a}' (\sigma)| =  | {\bf b}' (\sigma)| = 1~.
 \eeq
In the following, we will use this type of solutions in several different ways. First,
we will make use of this gauge to design simple initial conditions for the
numerical experiments that we will perform in the following section to check
the validity of the Nambu-Goto action. Moreover, we will also apply this
parametrization of the string to identify the (Nambu-Goto) initial data at 
any point in the evolution of the string and predict the subsequent evolution
according to this approximation. This will allow us to make a comparison 
of the string motion with the Nambu-Goto prediction at an arbitrary point during our
simulations. We describe this technique in more detail in Appendix \ref{appendix-NGreconstruction}.

\subsection{Testing the validity of the Nambu-Goto action}
\label{testing-NG}

The Nambu-Goto action is an approximate description of the
full dynamics of field theory domain walls. There are different types of corrections that
we can think of. First, there is the possible effect of radiation that we have
already discussed in the previous section. The results there indicate that
one can neglect these effects for low curvature domain walls.

Furthermore, it is also clear that the Nambu-Goto action is only 
strictly applicable to an infinitely thin wall. Therefore, if we want to 
apply it for field theory objects, there are additional
terms that need to be included in the effective action. In fact,
one can think of the NG action as the first term of an effective action
built as an expansion in powers of the ratio of the thickness of the string 
to its radius of curvature. This has been studied in some detail in 
several papers in connection to strings and domain walls \cite{Maeda:1987pd,
Gregory:1989gg,Gregory:1990pm,Anderson:1997ip}. 
Of course, there will also be instances where both these effects
could be important.

In the following sections, we will numerically evolve several 
initial conditions and study the validity of the Nambu-Goto action
and the circumstances where corrections play a significant
role.

\subsubsection{Colliding wiggles on a straight domain wall}

Here we will closely follow the setup studied in the case of local cosmic
strings in \cite{Olum:1999sg} and discuss the collision of wiggles
on a straight domain wall. In order to accurately compare the evolution
of the soliton with the NG prediction, we need to start with an initial
condition that is close enough to an exact solution of the full field theory
equations. Otherwise, the possible deviations from the NG dynamics
could just be due to the lack of precision in the initial configuration.

In this case, it is actually possible, using the results presented 
in \cite{Vachaspati:1990sk}, to set an initial condition that is
extremely close to an exact solution of the non-linear scalar field theory.
Let us describe how to construct this solution in detail.

Let's suppose once more that the domain wall is extended along the 
$y$ direction, and consider a field configuration of the form
\beq
\label{eq:psi plus psi minus}
\phi(x,y,t) = \phi_K \left[x - \psi_{+}(y + t) -  \psi_{-}(y + t) \, \right],
\eeq
where $\phi_K(x)$ is the domain wall profile and $\psi_{+}(y + t)$ and $\psi_{-}(y - t)$ 
represent wiggles propagating in opposite directions on the wall.
Using the results of \cite{Vachaspati:1990sk}, it is clear that
this is in fact a configuration that can be made to be
arbitrarily close to an exact solution initially provided that the two
functions $\psi_{+}(y + t)$ and $\psi_{-}(y - t)$ do not have significant 
overlap. For small amplitudes, these propagating wiggles correspond to the linear zero mode excitations of the type
presented in (\ref{eq:zm}).

So, what should we expect from these collisions? Let us first consider
the situation where the functions describing these wiggles are very mildly
curved, so even if we overlap them there would never be a point of large 
curvature for the domain wall. In this case, we will expect the NG
prediction to be accurate. Furthermore, looking at the expression in 
Eq. (\ref{NG-sol}) it is clear how to construct the NG initial data for this
configurations. Once we have the information for the functions ${\bf a}(\sigma)$ and ${\bf b}(\sigma)$
we can easily obtain the NG result for later times. This is all we need
to compare the field theory result, the one we obtain evolving the
complete field theory equations in a $(2+1)d$ lattice, and the 
NG predictions.

At small amplitudes for the displacement of the wiggles, one 
would expect them to propagate at the speed of light almost
unaffected by the presence of the other one. This behaviour
at small amplitudes can be explained by the fact, discussed earlier,
that wiggles on the string are perturbatively described by
zero modes at a linearized level.

However, the NG action is non-linear, as we stressed earlier.
Therefore, we would like to explore the situations where we can
probe these non-linearities. In other words, the NG action is a 
resumation of an effective action for all orders in first derivatives
of the displacements from the straight domain wall, so in order
to check this special structure of the theory one should 
go beyond linear order. 

On the other hand, the generic solution in the conformal
gauge given by Eq. (\ref{NG-sol}) seems to indicate that arbitrary wiggles colliding on 
a straight domain wall would just pass through one another
unaffected in their shape, so it is not clear what the non-linearities
do even in the large amplitude case.

The resolution of this puzzle is best explained in the static
gauge. In this case, one chooses the worldsheet coordinates to be
aligned with the straight wall, namely, we take
\beq
X^0 = t \,,~~~~~ X^1 = y~.
\eeq
The NG action in this case can be written as
\beq
 \label{NG-non-standard-gauge}
 S_{SG} = - \mu \int{dt ~dy \sqrt{1 + \psi'^2 - \dot \psi^2}}\,.
 \eeq
In this form, this is the action for a scalar field $\psi(y,t)$ living on a $2d$
worldsheet with a Born-Infeld type Lagrangian. The equation of motion from this action becomes
 \beq
 \label{eom-SG}
 \partial_t \left( \frac{\dot \psi}{\sqrt{1+\psi'^2 - \dot \psi^2}} \right) -\partial_y \left( \frac{\psi'}{\sqrt{1+\psi'^2 - \dot \psi^2}} \right) = 0~.
 \eeq
It is clear that a particular set of solutions of these equations is given
by waves moving at the speed of light along a straight domain wall. In other words, 
functions of the form $\psi(y\pm t)$ would be solutions of this equation. In this
way we recover the Vachaspati-Vachaspati configurations  \cite{Vachaspati:1990sk} within the
Nambu-Goto description. However, the equation is clearly non-linear and only when we have small amplitudes we 
will approximately recover the linear wave equation for a generic configuration
$\psi(y, t)$.

It is easy to see that the effect of the non-linear terms is only to create
a delay in the wiggles as they pass through one another. This delay can be
computed and it turns out to be proportional to the extra energy on the string
due to the existence of each of these wiggles. This is not surprising since the
difference between the parameter $\sigma$ in the NG solutions in the
conformal gauge and the space coordinate $y$ which parametrizes our
solutions in this other gauge is indeed given by this amount of 
energy\footnote{To be more precise, we can think about a small wiggle first propagating on a long straight string. In
this case there is no distinction between the coordinate $y$ and the amount of
energy that the wiggle traverses, $\sigma$. However, as the wiggle encounters
the oppositely moving large excitation, these two quantities are not the same. The NG dynamics
tells us that this small wiggle will travel at the speed of light in $(t, \sigma)$ space
and not in $(t,y)$ space. The difference between these two quantities is exactly the
extra energy of the wiggle with respect to the straight string. Hence, there will be a
delay in the position of the little wiggle after the collision between its actual position
and the value of the $y$ coordinate one would (erroneously) predict assuming that
it had propagated at the speed of light in the $(t,y)$ coordinates.}.

Thinking in terms of the conformal gauge solution helps us understand
this scattering process. Since the theory is integrable, the wiggles
can not change their shape or produce $(1+1)d$ radiation in the
collision. Therefore, it is obvious that this is the only possible outcome of this
interaction between the asymptotic waves. This type of behaviour is 
reminiscent of the interaction of solitons in integrable models like the Sine-Gordon equation \cite{Rajaraman:1982is}.

\begin{figure}[h!]
\includegraphics[width=17cm]{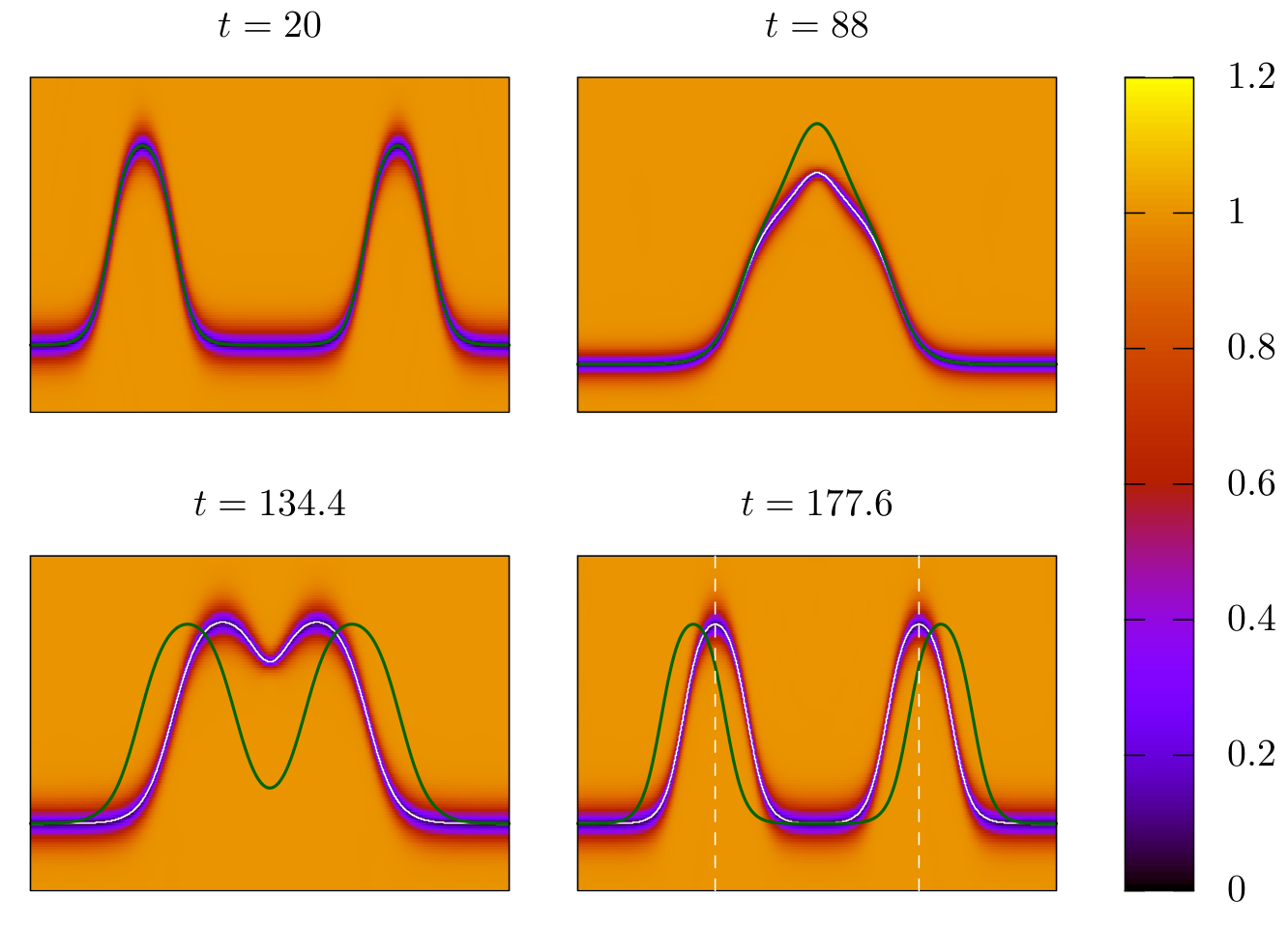}
\caption{Field theory simulation of the collision of low curvature wiggles propagating on a straight 
domain wall string. We show the absolute value of the field, $\phi/\eta$,  in the 2-dimensional simulation
space (note that according to our prescription in the main text the straight domain wall string is 
extended along the $y$ direction, which here corresponds to the
horizontal axis). The white curve represents the position of the domain wall string 
according to the NG approximation, while the green curve corresponds to the solution 
of the linear wave equation. Finally, the dashed vertical lines represent the predicted position of the
peak of the wiggles taking into account their time delay due to the non-linearities of the NG equation of
motion. The time labels are displayed in units of $\left(\sqrt{\lambda}\eta\right)^{-1}$, and the ranges of $x$ and $y$ in these units are the following: $y\in\left[-150,150\right]$ and $x\in\left[-5,20\right]$ for the $t=20$ and $t=177.6$ panels, $y\in\left[-100,100\right]$ and $x\in\left[-5,30\right]$ for the $t=88$ panel, and $y\in\left[-100,100\right]$ and $x\in\left[-5,20\right]$ for the $t=134.4$ panel. }
\label{fig:nonlinearity NG}
\end{figure}

We have checked these predictions of the NG approximation by performing 
several different numerical simulations in field theory and comparing them with
the results expected from the NG dynamics. In order to do this, we have developed
a method of reconstruction of the NG evolution from the position and velocity
of the domain wall string in field theory. See the Appendix \ref{appendix-NGreconstruction} 
for the details of this reconstruction method.

We started by considering the collision of two wiggles designed in such a way that their NG evolution would
lead to a domain wall configuration with a small curvature everywhere.
Our simulations show that the NG prediction in this case is indeed
perfect. The reconstructed NG energy stays constant and the position of the string obtained directly from the simulation agrees precisely with the one predicted in NG (see Fig. \ref{fig:nonlinearity NG}).

Furthermore, we also plot the position that the wiggles would have if the solution
was just the wave equation. We can see in Fig. \ref{fig:nonlinearity NG} a comparison between
these different situations, where we can easily appreciate the delay in the field
theory simulation due to the non-linear NG interactions. This is also in 
perfect quantitative agreement with the NG description. We also plot in this figure
the predicted position of the peak of the wiggles assuming the delay is given by the
extra energy of the oppositely moving wiggle with respect to the straight string. We note
the agreement is again perfect. We have also performed this type of experiments
with asymmetric wiggles and found, in all cases, a perfect agreement between the
numerical results and the analytic NG description.

\begin{figure}[h!]
\includegraphics[width=13cm]{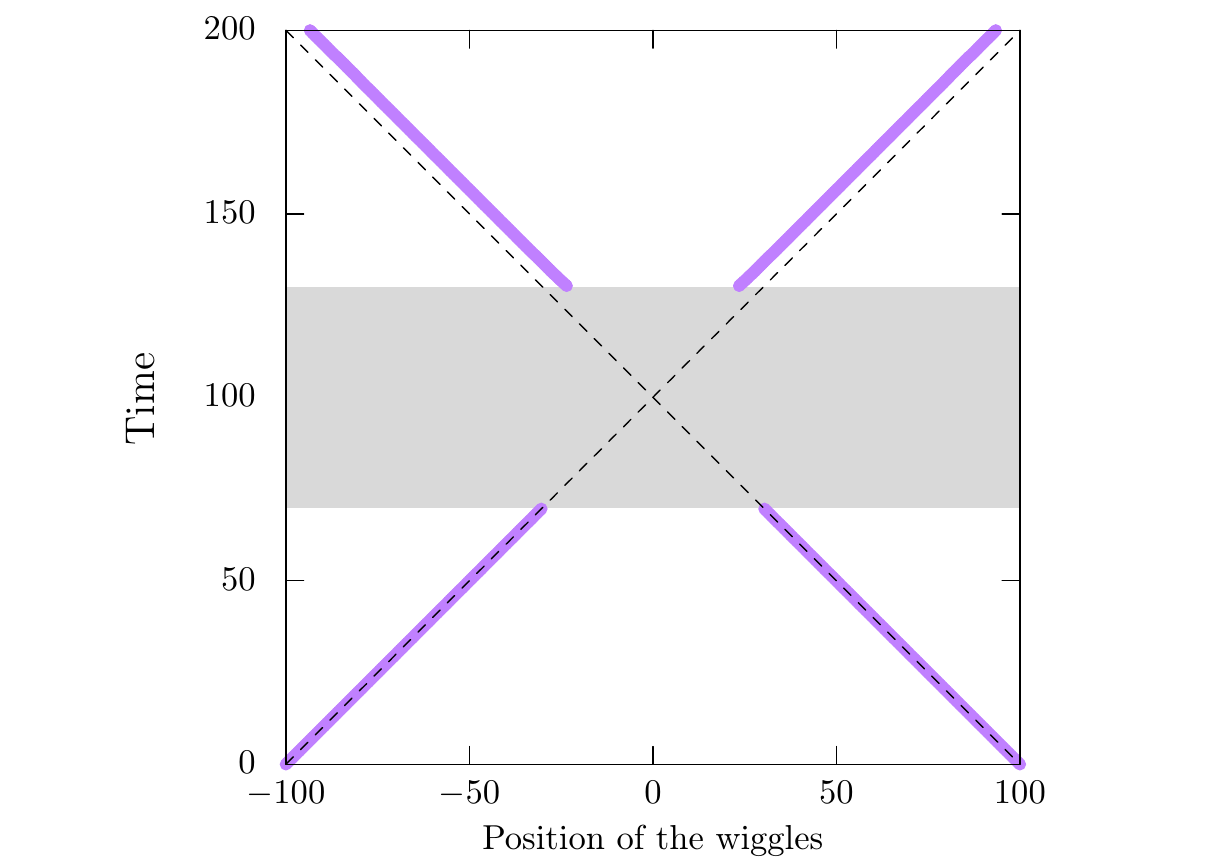}
\caption{Spacetime diagram for the position of the wiggles propagating on the straight domain wall string
obtained directly from the field theory simulation (in purple) and for speed of light propagation (dashed line). 
The shaded grey region indicates the time during which the wiggles overlap.
} 
\label{fig:spacetime diagram}
\end{figure}

Finally, we visualize this delay  by plotting in Fig. \ref{fig:spacetime diagram} the position 
of the center of the wiggles in a $(y,t)$ spacetime diagram.

We have also let the wiggles interact several times by
imposing periodic boundary conditions on the $y$ axis. The results
continue to be in perfect agreement with NG even for a large number
of collisions.

\begin{figure}[h!]
\includegraphics[width=13cm]{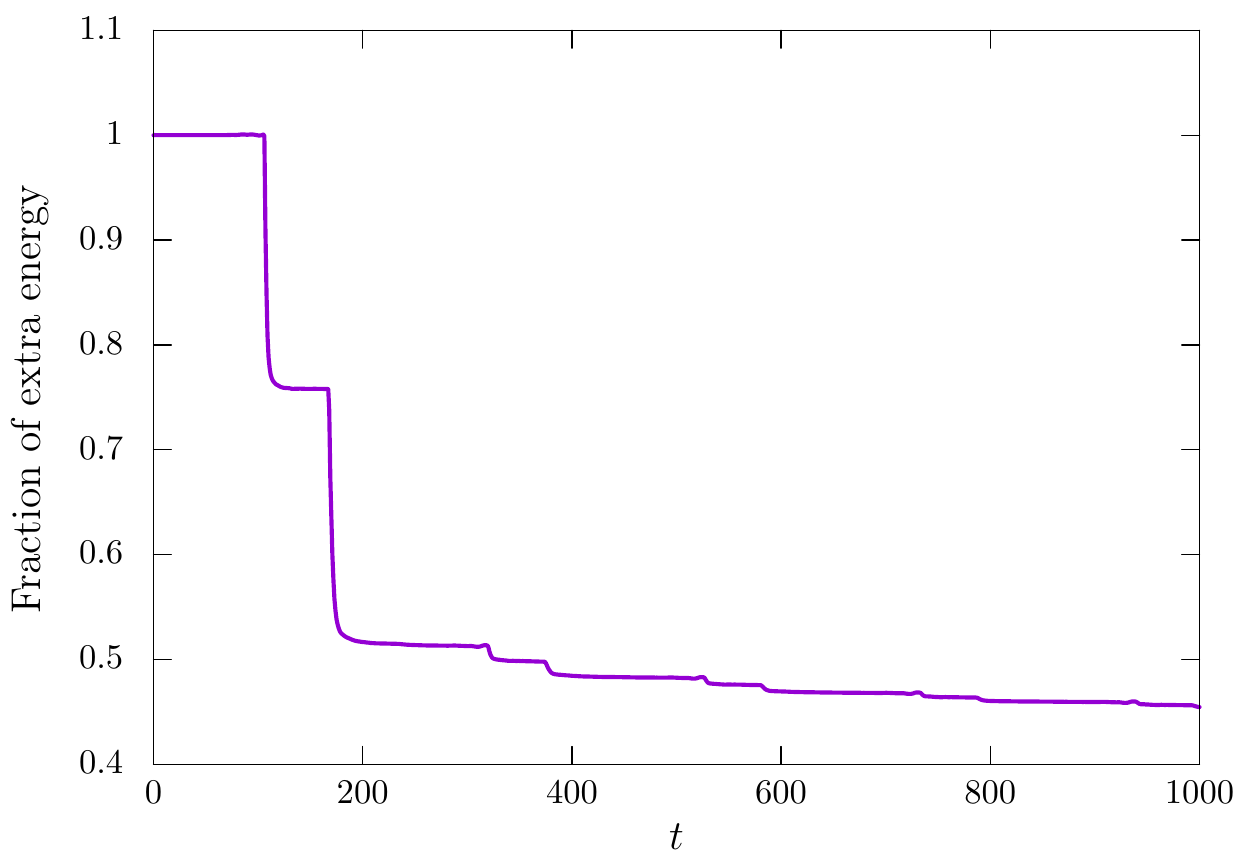}
\caption{Fraction of extra energy as a function of time, displayed in units of 
$\left(\sqrt{\lambda}\eta\right)^{-1}$, for high curvature wiggles colliding on 
the wall. ``Extra energy'' means total energy inside the box minus energy of 
the straight domain wall, or in other words, the energy on the wiggles on the straight
domain wall. In this case, the maximum ratio of string thickness to radius of 
curvature is approximately $\kappa\approx 0.44$ at the moment of the collision.} 
\label{fig:energy high curvature}
\end{figure}

Finally, we have also explored the scattering of wiggles that lead to
the formation of high curvature regions. We show in Fig. \ref{fig:energy high curvature} the amount 
of energy in the box in our numerical
simulations. We see the different steps where the energy is depleted
by the bursts of radiation from the high curvature regions. This is
very similar behaviour to the one presented in  \cite{Olum:1999sg}.
After the collision, the domain wall behaves again like a NG soliton
with wiggles of different shape.

In summary, we have checked explicitly that the non-linear behaviour of
the NG action reproduces perfectly the solution seen solving the full field theory equations
in the case of small curvatures for the domain wall.

\subsubsection{Standing waves}

In this section, we compare the evolution of a zero mode excitation in the form of a standing wave
in the NG description with the actual solution from field theory.
Similarly to what we use in section IIIA, we will take our initial 
conditions for the standing wave in field theory to be the ones described by 
Eq. (\ref{FT-for-NG}). We stress that these are not exact initial
conditions, but only approximate ones. However, following the arguments in Appendix \ref{appendix-ansatz for NG dynamics},
we will consider them a good enough approximation for our purposes. Furthermore, choosing the amplitude and the frequency 
of the standing wave with some care (see Eq. (\ref{def-curvature})), 
we can study the behaviour of solutions
within the small curvature regime where the NG action should be
valid. 

We analyze different values of the frequency of oscillation, some
below the threshold of the massive particles in the bulk and 
some above. The results show that for frequencies below the
threshold the amount of radiation is negligible and the position of the wall
stays within the NG solution for extremely long periods of time. 
We compare both the position of the domain wall with the 
NG prediction as well as the energy from the NG reconstruction.
Both measures indicate a very good agreement with the
NG description. For higher frequencies, the emission of radiation
reduces the NG energy in time in agreement with the
prediction from the analytic estimates of Appendix \ref{an_zero}.

For larger amplitudes but frequencies below the threshold of radiation,
the agreement with NG becomes less precise. The reconstructed NG energy is 
still more or less constant, but the position of the domain wall
is slightly different from the NG dynamics. This may be an indication
of the effect of higher order terms in the effective action missing
in the NG dynamics. However, this is hard to quantify since increasing
the amplitude some more one starts seeing the effects
of non-perturbative radiation so it is difficult to disentangle
both effects.\\

Our results in this section confirm the expectation that the NG
action describes very accurately the evolution of domain wall
strings without any internal excitation. This agreement ceases
to be perfect only in the presence of high frequency oscillating modes
where some radiation is possible. This was also expected since 
at those frequencies one can start exciting the massive degrees
of freedom that could propagate in the bulk carrying away part
of the energy of the domain wall string. Finally, regions of high
curvature also lead to deviations from NG and the emission
of energy by a non-perturbative process; a sort of high
curvature annihilation.

\section{Effective theory including the internal mode}

As we described earlier, the presence of excitations localized on the domain
wall could add a substantial amount of energy to the configuration. This suggests
that one could change the dynamical properties of the soliton by this excitation
process. This is somewhat similar to what happens with superconducting cosmic
strings, where the current on the string core effectively modifies
the equation of state of the string \cite{Witten:1984eb,Davis:1988ij}.

A simple strategy to quantify this effect would be
to analyse the energy momentum tensor for the excited domain wall as a function of the
amplitude of the internal mode. This will give us an idea of the kind of
change we could expect in its dynamics for a given amount of extra energy
in the excitation.

In order to do that, we consider the simple field configuration of a homogeneously
excited straight wall of the form
\beq
\phi(x,y,t) = \phi_K(x) + A(t) \times f_1(x)~,
\eeq
and insert it in the expression for the energy momentum tensor
of the scalar field. After integrating over the transverse $x$ direction,
one obtains, at a quadratic order in the amplitude of the excitation, the 
following expression for the energy per unit length and tension of the domain
wall string,
\beq
\mu_A= \mu \left(1+ \frac{9}{8\sqrt{2}}\frac{\sqrt{\lambda}}{\eta}\hat A^2 + ... \right)
\eeq
and
\beq
\label{oscillatingtension}
\tau_A(t) = - \mu \left(1 +  \frac{9}{8\sqrt{2}}\frac{\sqrt{\lambda}}{\eta} \hat A^2 \cos(2\omega_s t)+ ... \right)\,,
\eeq
where we have assumed that the evolution of the shape mode is well 
approximated at quadratic order by a periodic function of the form\footnote{This assumption
receives corrections when one includes higher order terms, but we will
limit ourselves with these leading order terms for now.} $A(t) = \hat A \cos( \omega_s t)$.

One can obtain this type of energy momentum tensor postulating the
existence of a domain wall with an extra scalar degree of freedom
living on its worldsheet. The simplest effective action of this kind is given by
\beq
\label{effective-action}
S = \int{d^2\xi \sqrt{-\gamma} \left[- \mu + \frac{1}{2} \gamma^{\alpha \beta} \partial_{\alpha} \theta \partial_{\beta} \theta - \frac{1}{2} m^2_{\theta}\theta^2  \right] }\,,
\eeq
where as before $\gamma$ parametrizes the induced metric on the worldsheet and
the scalar field $\theta(\xi^a)$ characterizes the amplitude of the internal mode excitation
provided that we choose $m_{\theta}=\frac{\sqrt{3}}{2} m$.

At a linear order, this theory describes the free degrees of freedom propagating on the
straight domain wall, namely, the massless Goldstone modes and the
massive internal mode. These are just the fluctuation modes we identified in the previous section. 
Taking it beyond the linear order, the theory describes a non-trivial interaction between 
these modes which could lead to interesting effects.

On the other hand, it is clear that the massive mode in this $1+1$ dimensional theory introduces
a new time scale associated with its oscillating period $\tau_m \sim 1/m_{\theta}$.
For times much larger than this, the average effect of the excitation on a 
straight domain wall is to increase its energy density
but does not modify its tension. For a uniformly excited wall, this would seem to 
decrease the speed of propagation of the transverse excitations of the string.
Thinking about a closed loop, this would also seem to slow down the 
collapse of the loop.

However, it is important to remember that this effective action disregards any radiative
effect. This means that in order to use it we should only consider situations where 
the excitations are not large enough to create a substantial amount of radiation in the
period of time that we are interested in. Otherwise, the real dynamics of the domain wall in 
field theory would be very different from the one inferred from this action. 

As we described in detail in previous sections, the presence of an excited
domain wall leads to radiation on a time scale that is inversely proportional to
the square of the amplitude of the excitation. This means that if we are interested in studying
the implications of this effective action over long periods of time we should
only consider the case with small amplitude values for $\theta$. However,
in this case, the possible deviations from the Nambu-Goto 
action would therefore be small. One could try to incorporate these
effects into this effective action, but it is not clear to us how to do this
in an accurate way.

Alternatively, one could always study the effects of these excitations
by using field theory simulations in the lattice. We have done one such
simulation of a collapsing domain wall ring initially at rest and excited in a uniform
way. The results are presented in Appendix \ref{appendix-ring}, where we show how 
the evolution of the domain wall loop is slowed down by the presence
of the extra energy associated to the scalar field perturbation. This is
in agreement with the results obtained from the effective action given
by Eq. (\ref{effective-action}). However, a detailed comparison of this type of 
configuration with the results obtained from this
effective action is difficult due to the presence of extra radiation for
large values of the amplitude of the excited state. \\

\subsection{Parametric resonance within the effective action description}

Let us conclude this section by indicating another important 
effect that is already hinted in our effective action approach. As 
one can see in Eq. (\ref{oscillatingtension}), the value of
the tension of the domain wall oscillates in the presence of 
internal modes. This suggests that one could have parametric
resonance effects that would stimulate the appearance of 
transverse excitations on the position of the wall. This is
the analogue effect to the non-relativistic string resonance already
discussed a long time ago in \cite{Rayleigh}. Indeed, one can
easily show this behaviour by looking at the equation of motion
for the position of the domain wall string from our effective action, namely the
equation

\beq
\partial_a \left[ \sqrt{-\gamma} (\mu \gamma^{ab} + T^{ab}) \partial_b X^{\mu}\right] =0\,,
\eeq
where $T^{ab}$ denotes the $(1+1)d$ energy momentum tensor
associated with the scalar field describing the bound state amplitude, the scalar field $\theta$.
Assuming the lowest order solution for the scalar field equation of 
motion, we take $\theta(t) = \theta_0 \cos(w_s t)$. 

Let us now consider a string extended along the $y$ direction that
is parametrized in the form $ X^{\mu} (t,y) = (t, \psi(t,y), y)$. And
for simplicity let us take the displacement of the string to be
$\psi(t,y)= D(t) \cos(w_0 y)$. Within this ansatz, the equation for
the amplitude of the string displacement is given by

\beq
\ddot D(t) + w_0^2 \left( 1 - \frac{\dot \theta^2}{\mu} \right) D(t)=0\,,
\eeq
which for our oscillating scalar field transforms into a form of the
Mathieu equation. This shows that this type of internal mode excitation
can create a parametric instability in the position of the domain
wall of a form of a standing wave of frequency equal to the
bound state frequency, namely, $w_0=w_s$.

In the following section we will explore the presence
of these instabilities in full detail at the non-linear level
with the help of our numerical simulations. In fact, we will
see that these are not the only source of parametric
resonance effects present in field theory.

\section{Parametric Resonances from Field Theory simulations}

\subsection{Dimensionless variables}

From now on, to make the comparison to our numerical simulations more direct,
we will work with the following set of dimensionless variables:
\begin{equation}
\tilde{\phi}=\phi/\eta,\,\,\,\,\tilde{t}=\sqrt{\lambda}\eta t,\,\,\,\,\tilde{x}=\sqrt{\lambda}\eta x,\,\,\,\,\tilde{y}=\sqrt{\lambda}\eta y\,.
\label{eq:dimensionless}
\end{equation}
With this rescaling, the action reads
\begin{equation}
S=\frac{\eta}{\sqrt{\lambda}}\int d^{3}\tilde{x}\left[\frac{1}{2}\partial_{\mu}\tilde{\phi}\partial^{\mu}\tilde{\phi}-\frac{1}{4}\left(\tilde{\phi}^{2}-1\right)^{2}\right]\,,
\label{eq:rescaled action}
\end{equation}
where the partial derivatives are now with respect to the dimensionless spacetime 
coordinates. The mass of small fluctuations about the vacuum, the energy per unit 
length of the static domain wall solution and the angular frequency of the homogeneous 
shape mode are now given by $\tilde{m}=\frac{m}{\sqrt{\lambda}\eta}=\sqrt{2}$, 
$\tilde{\mu}=\frac{\mu}{\sqrt{\lambda}\eta^{3}}=\sqrt{8/9}$ and 
$\tilde{\omega}_{s}=\frac{\omega_{s}}{\sqrt{\lambda}\eta}=\sqrt{3/2}$, respectively. Moreover, 
the amplitude of the shape mode will be rescaled as $\tilde{\hat{A}}=\lambda^{1/4}\hat{A}/\sqrt{\eta}$, which is still dimensionless. 

In the following we will drop the tildes for simplicity, but all variables will refer to these new dimensionless quantities. 

\subsection{A parametric instability for a homogeneous internal excitation}
\label{subsec:nonlinear}

As we pointed out in the previous section, the perturbation of a domain wall by 
an internal mode can lead to the subsequent excitation of a particular zero mode. This
type of parametrically resonant excitations of the transverse position of the string
by perturbations on its tension have been known and studied for a long time. In fact, 
this is one of the first examples of parametric resonance events in the history of 
physics \cite{Rayleigh}.

The simplest way to see this phenomenon in our field theory context is to 
assume a particular ansatz for the perturbations around the static
straight domain wall of the form
\beq
\label{eq:dw plus shape plus zero}
\phi(t,x,y) = \phi_K(x) + A(t) f_1(x) + D(t) f_0 (x)\cos (\omega_{0} y)~,
\eeq
where $f_0(x)$ and  $f_1(x)$ are given by the wave functions found in
Section \ref{spectrum}.
We will consider the string to be in a box with periodic boundary
conditions along the $y$ direction. This is, of course, a particular truncation 
of the most generic type of ansatz we can write. In particular, we are disregarding 
any coupling of this configuration to any radiating mode. This would only be 
consistent with our findings in the previous sections if we assume that the 
amplitudes of the modes involved in this calculation are sufficiently small.

Plugging this expression in the equation of motion for the scalar field
and projecting the result onto the different mode functions of the perturbations,
we arrive to a set of coupled differential equations for the amplitudes of the
different modes\footnote{See the Appendix \ref{appendix-lagrangians} for a detailed explanation of this 
procedure.}:
\beq
\label{eq:reducedEqs1}
\ddot A(t) +\frac{3}{2} \left[ 1+ C_1 D^2(t) \right] A(t)+ 6 C_2 A^2(t) + \frac{3 C_1}{2} A^3(t) +  \frac{3 C_2}{2} D^2(t) = 0\,,
\eeq
\beq
\label{eq:reducedEqs2}
\ddot D(t) + \left[ \omega_{0}^2 + 6 C_2 A(t)+ 3C_1 A^2(t) \right] D(t)  + \frac{9 C_1}{4} D^3(t)  = 0\,,
\eeq
where the numerical coefficients come from the projection over the $x$ direction and
are given by $C_1 = 3\sqrt{2}/35$ and  $C_2 = 3\sqrt{3}\pi / (64 ~2^{3/4}) $.

Taking the limit of $D=0$ of this system of equations we
obtain the non-linear equation for the oscillation of the internal mode. Reducing the 
equation to the linear order we recover the result that the internal mode oscillates
with a frequency $\omega_s = \sqrt{3/2}$.

Let us now consider the equation for the amplitude of the zero mode. Taking
the lowest order approximation of this equation we arrive at
\beq
\ddot D(t) + \left[ \omega_{0}^2 + 6 C_2 A(t) \right] D(t) = 0~.
\eeq

Assuming that the shape mode is excited at a linear order, we can substitute
its amplitude in this equation with the expression $A(t) = \hat A \cos(\omega_s t)$.
Doing this substitution, we can identify that the resulting equation is nothing
but a particular form of the Mathieu equation \cite{MathieuEq}. This type of equation has
been extensively studied in the context of parametric resonance in different
physical phenomena. The results of these studies indicate that the solution
of this equation has several bands of instability centered around particular
values of the frequency $\omega_{0}$ where its amplitude would grow exponentially
with time \footnote{See the brief discussion in the Appendix \ref{matsec} of the most relevant aspects of
the Mathieu equation and its connection with our problem here.}. In our case, this translates 
to a region of instability around $\omega_{0} \approx \omega_s/2$.

Note that this is a different resonant frequency than the one obtained
from the effective action in the previous section. The reason for this
is clear: the resonant effect comes in this field theory language from
a linear term in the amplitude of the internal mode, while it is of quadratic
order in the effective action description. We will comment more on this
in the next section, but let us just mention that a quadratic driving resonant
term also appears in the field theory equations in Eq. (\ref{eq:reducedEqs2}).
However, as the amplitude of the internal mode is small, this term will not be the
dominant one here. In order to test this prediction, we perform a field
theory simulation in a lattice.

\begin{figure}[h!]
\includegraphics[width=17cm,height=7cm]{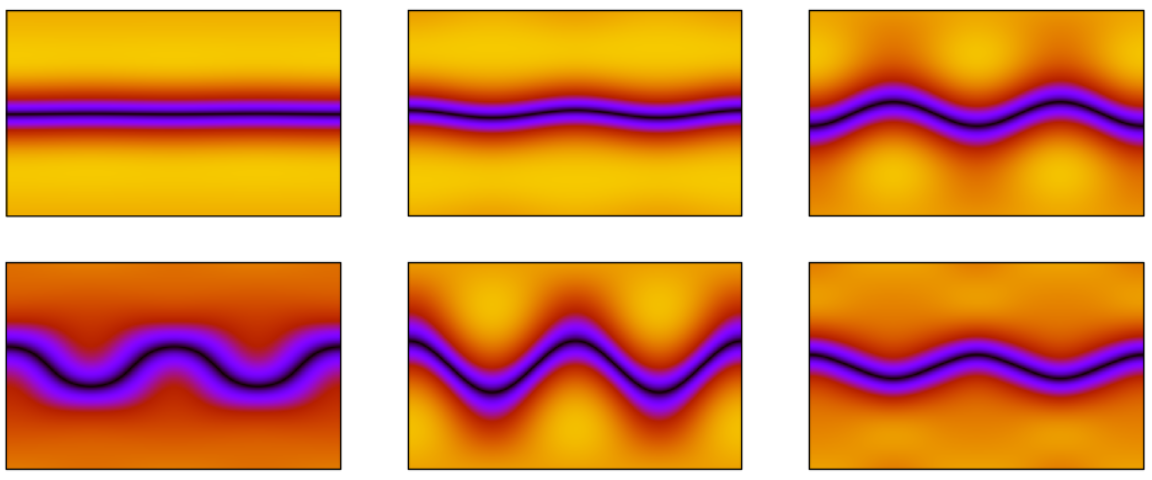}
\caption{Parametric resonance excitation of a standing wave for the position of the
domain wall string with frequency $\omega_{0}\approx\omega_{s}/2$ for a homogeneous initial
state with an internal mode excitation of frequency $\omega_s$. In these plots, $x$ ranges from $-4$ to $4$ (vertical axis), and $y$ from $-10$ to $10$ (horizontal axis).  } 
\label{fig:resonant zero mode}
\end{figure}

Following the previous calculation we will expect that the homogeneous initial
configuration of a domain wall with an excited internal state would resonantly
produce a transverse perturbation of the form of standing waves with 
frequency $\omega_{0} \approx \omega_s/2$. We show in 
Fig. \ref{fig:resonant zero mode} several snapshots of the position of the domain wall for this
type of configuration from our numerical full field theory simulations. For this example, we took the initial conditions to be
\begin{equation}
\phi\left(x,y,t=0\right)=\phi_{K}\left(x\right)+A(0)f_{1}(x)+D(0)f_{0}(x)\cos\left(\omega_{0}y\right)\,,
\label{eq:initial condition phi 6b}
\end{equation}
\begin{equation}
\dot{\phi}\left(x,y,t=0\right)=0\,,
\label{eq:initial condition pi 6b}
\end{equation}
with $\omega_{0}=2\pi/10$, $A(0)=0.5$ and $D(0)=0.02$.
It is 
clear from the images that a transverse modulation of the wall grows over time,
becoming really noticeable in agreement with the previous analytic considerations.

This exponential growth is eventually tamed by the higher order terms in the different
amplitudes of the modes that we have neglected in our discussion. In particular, as the zero mode amplitude 
grows, the one for the shape mode decreases. This is
basically a process of energy transfer between these modes. After a while,
the non-linear source term proportional to $D^{2}(t)$ in the equation for
the amplitude of the shape mode becomes important. This term drives the 
amplitude for the shape mode back up to its original value, depleting, in turn, 
the amplitude of the zero mode. This completes the full cycle and brings the system to the initial state.

\begin{figure}[h!]
\includegraphics[width=13cm]{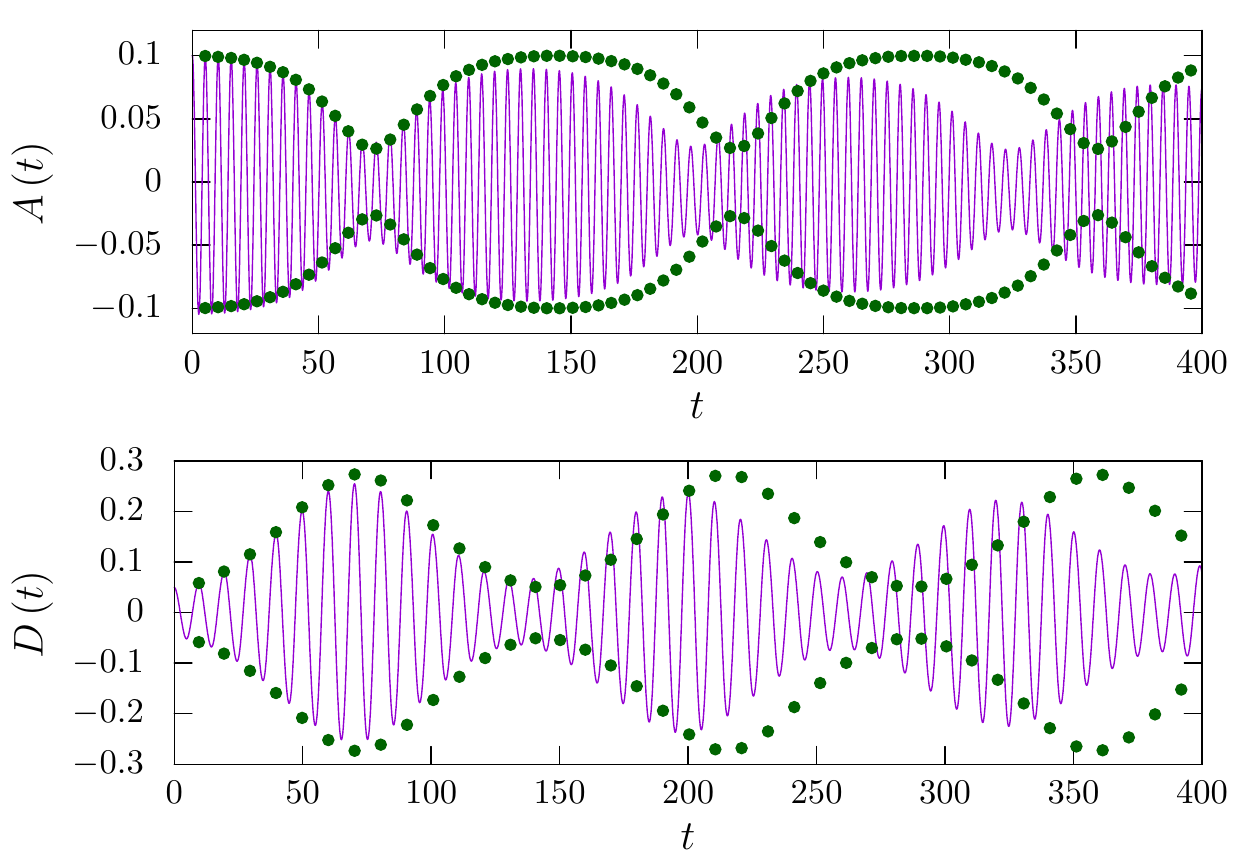}
\caption{Amplitude of the shape mode (top panel) and the resonant zero mode (bottom panel) 
as a function of time. The values read from the simulation are shown in purple, and the green dots correspond 
to the solution of the coupled equations (\ref{eq:reducedEqs1}) and (\ref{eq:reducedEqs2}). In the latter case, we 
only show the envelope in order to make the comparison easier.} 
\label{fig:shape and zero coupled eqs}
\end{figure}

We have performed several numerical simulations for an initial homogeneous
state of an excited domain wall. Using the orthogonality condition of the modes
that describe the linear perturbations of the static domain wall solution, we can
extract directly from the simulation the values of the amplitudes of these modes
as a function of time. We have done this for a particular run and compared the results
with the numerical solution of the reduced set of Eqs (\ref{eq:reducedEqs1}) and (\ref{eq:reducedEqs2}). The results are shown
in Fig. \ref{fig:shape and zero coupled eqs}. Although the first pulse is predicted almost 
perfectly by the coupled equations, the subsequent ones deviate appreciably presumably
due to the presence of radiation being emitted in the process. Indeed, we do observe this radiation in our simulations.

\subsection{A standing wave on the internal mode}

\begin{figure}[h!]
\includegraphics[width=15cm]{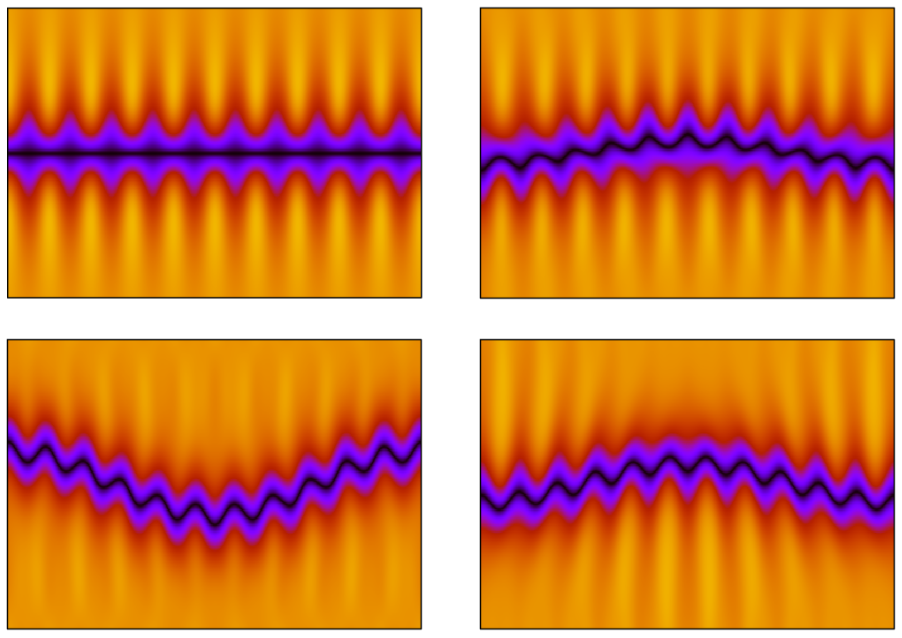}
\caption{Standing wave in the internal mode decaying into a standing wave on the position of the domain wall
string with two different frequencies. In these plots, $x$ ranges from $-5$ to $5$ (vertical axis), and $y$ from $-17$ to $17$ (horizontal axis).} 
\label{fig:shape mode standing wave}
\end{figure}

Following the discussion in the previous section, we would like to explore now the 
possibility of other resonant phenomena in more generic situations. In particular, we will consider
the case where the internal mode has the form of a standing wave. In this case, 
we can show that the initially straight domain wall develops an instability towards
the formation of a standing wave with a couple of different wavelengths (see the 
snapshots in Fig. \ref{fig:shape mode standing wave}).

We can explain this behaviour by taking the ansatz
\beq
\label{eq:dw plus shape plus zero plus zero}
\phi(t,x,y) = \phi_K(x) + A(t) f_1(x)  \cos (k_s y) + D_1(t) f_0 (x) \cos (k_1 y) + D_2 (t) f_0 (x)\cos (k_2 y)~.
\eeq

The resultant equations of motion at the lowest order for the time dependent amplitudes of the different
modes turn out to be, after the projection on the $x$ and $y$ directions,
\bea
\ddot A(t) +\left( \frac{3}{2}+ k_s^2 \right) A(t)+  3C_2 D_1(t) D_2(t) &=& 0~,\\
\ddot D_1(t) +  k_1^2 D_1(t)  + 3C_2 A(t) D_2(t)  &=& 0~,\\
\ddot D_2(t) + k_2^2  D_2(t)  + 3C_2 A(t) D_1(t)  &=& 0~,
\eea
where $C_3 = 9\sqrt{3}\pi / (64 ~2^{3/4})$. These are the correct equations for the amplitudes as long as 
the following three conditions are met: (1) $k_{1}\neq k_{s}/2$, (2) $k_{2}\neq k_{s}/2$ and
(3) $k_{1}+k_{2}=k_{s}$ or $|k_{1}-k_{2}|=k_{s}$. This kind of system of coupled equations has been 
discussed before in the literature in connection with mechanical systems (see for example \cite{METTLER1967169}).
The results in these papers indicate that one can find solutions where a simultaneous resonant 
amplification of two different modes is possible provided that their wavenumbers satisfy the condition
\beq
k_{1}+k_{2}=\Omega_{s}=\sqrt{\frac{3}{2}+k_{s}^2} ~.
\eeq
One can in fact find bands of instability quite similar in nature to the ones in the
Mathieu equations, so this condition does not have to be completely
sharp. If we also take into account condition (3) above, we should expect amplification of two zero modes 
with frequencies $k_{1}=\left(k_{s}+\Omega_{s}\right)/2$ and $k_{2}=\left(-k_{s}+\Omega_{s}\right)/2$. This is 
exactly what we have found in our numerical simulations of the excited
domain wall in a box\footnote{There is an added difficulty in our setup due to the
periodic boundary conditions along the $y$ direction. This adds the extra condition
to the appereance of the instability due to the quantization of the wavenumbers 
along that direction.}.
For the simulation shown in Fig. \ref{fig:shape mode standing wave}, we took the following initial conditions:
\begin{equation}
\phi\left(x,y,t=0\right)=\phi_{K}\left(x\right)+A(0)f_{1}(x)\cos\left(k_{s}y\right)\,,
\label{eq:initial condition phi 6c}
\end{equation}
\begin{equation}
\dot{\phi}\left(x,y,t=0\right)=0\,,
\label{eq:initial condition pi 6c}
\end{equation}
with $k_{s}=10\pi/17$, $A(0)=0.4$.

\subsection{An effective action including higher order interactions}

As we mentioned already, the parametric resonance predicted using the
effective action in Eq. (\ref{effective-action}) yields a different result for the resonant zero
mode frequency of excitation than the one found in field theory. This 
discrepancy can be traced to the relevant oscillatory term in the
associated Mathieu equations. In the effective theory description, 
that term is quadratic on the amplitude of the internal mode. However,
in the field theory account of this effect, the relevant term is linear.
This explains the doubling of the resonant frequency between both
models.

An interesting question that we might ask is whether we can 
supplement the effective action given by Eq. (\ref{effective-action}) in such a way that it
can accommodate the coupling between the modes that we have uncovered
in the previous section in the field theory simulations.

One possibility could be the presence of terms of the form
\beq
S_{\text{curvature}}= \beta \int{ d^2\xi \sqrt{-\gamma}~ \theta(\xi) \cal{R}}\,, 
\label{eq:coupling theta ricci}
\eeq
where the scalar field controlling the internal mode amplitude is
coupled to the spacetime curvature of the domain wall string worldsheet
($\cal{R}$). Note that $\cal{R}$ depends on derivatives of the position
of the domain wall string $X^{\mu}$. Therefore, this term describes a coupling between
the internal mode and the Goldstone modes on the string. Terms of this form, 
without the coupling to the worldsheet scalar, have been proposed already
in the context of higher order corrections to NG \cite{Maeda:1987pd,
Gregory:1989gg,Gregory:1990pm,Anderson:1997ip}. The interesting
point about this addition to the effective action is that looking at the
linearized equations of motion one identifies the appearance of a
parametric resonance with the correct frequency (we show this in Appendix \ref{R-coupling}).
Furthermore, comparing the result with the field theory description, one can 
infer the value of the coefficient $\beta$ in front of this term.
Of course, this is not a proof that this is the correct term to the
effective action, just that we can accommodate this particular
coupling between the Goldstone mode and the internal mode
with this term \footnote{It is reasonable to assume that there
should also be a term similar to this one without the $\theta$ 
dependence in the effective action. However, this is not very
relevant for our discussion here.}.
We leave the question of the relevance of these terms to be 
studied in a future publication.

\begin{figure}[h!]
\includegraphics[width=14cm]{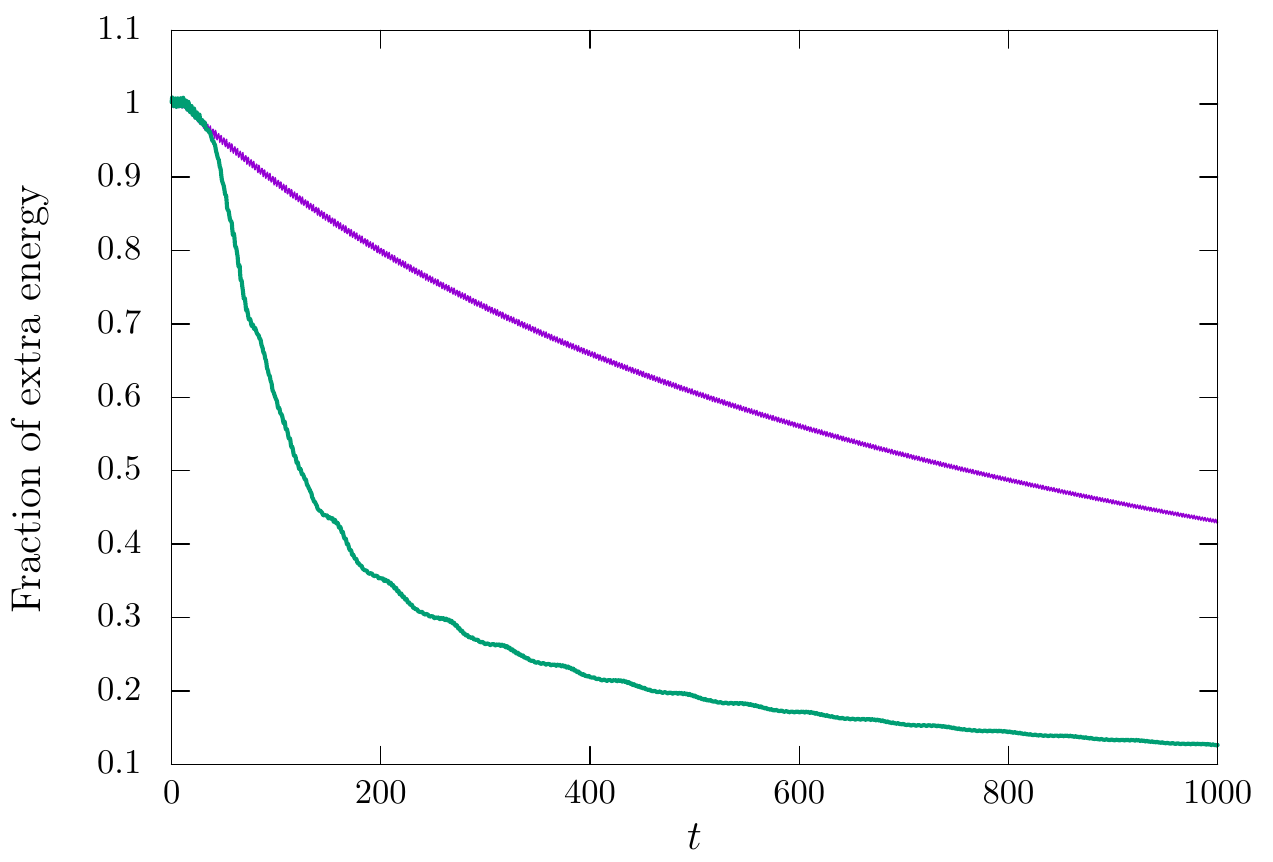}
\caption{Fraction of extra energy inside the box as a function of time in a $1+1$ 
simulation (purple curve) and in a $2+1$ simulation where the resonance 
occurs (green curve). ``Extra energy'' means additional energy with respect 
to the straight domain wall solution. The initial amplitude of the shape mode 
is, in this case, 0.3.}
\label{fig:new time scale radiation}
\end{figure}

\subsection{New time scale of radiation}

The discussion in the previous sections shows that the presence of excitations of the
internal mode triggers the subsequent resonant excitation of Goldstone modes. This
non-linear process can potentially also lead to radiation from the domain walls. Here
we show that this is indeed what happens. We show in Fig. \ref{fig:new time scale radiation} 
the energy per unit length of an initially excited state in $1+1$ dimensions and compare 
it with the same case within a $2d$ box. The configuration in $1+1$ dimensions is, of course,
prevented from exciting any longitudinal mode by construction and decays exactly
according to the rate given by Eq. (\ref{eq:manton law}) in the Appendix \ref{an_shape}. Allowing 
for the domain wall to resonantly oscillate along the $y$ direction, as in the simulation in
the $2d$ box, induces a higher rate of radiation. 

This demonstrates that the non-linear interaction between the internal mode and
the Goldstone modes leads to a quicker emission of the extra energy in the form
of the internal excitation. In other words, this interaction reduces the lifetime of
the internal excitations.

\section{Dynamical excitation of the internal modes}

One of the interesting points about the existence of internal
excitations in these models is their relatively long lifetime. We have
shown that for small amplitudes these excitations could survive
for a time much larger than their natural time scale given by
the thickness of the soliton. This makes these 
modes potentially relevant for numerical simulations where the
time scale of the runs is shorter than the lifetime of these excitations.

 \begin{figure}[htb]
\includegraphics[width=16cm, height=4cm]{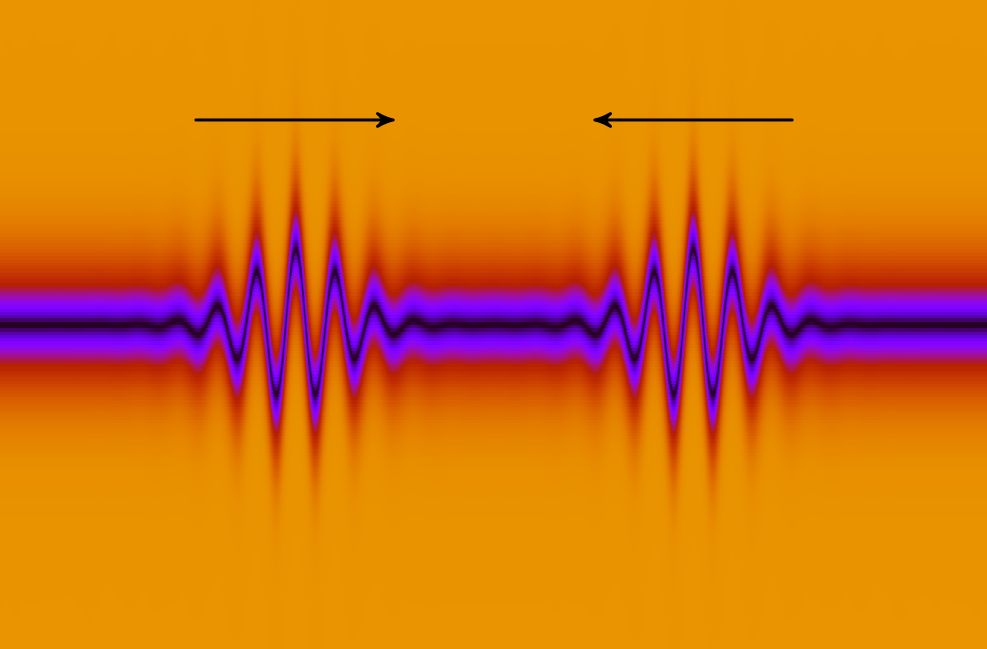}
\includegraphics[width=16cm, height=4cm]{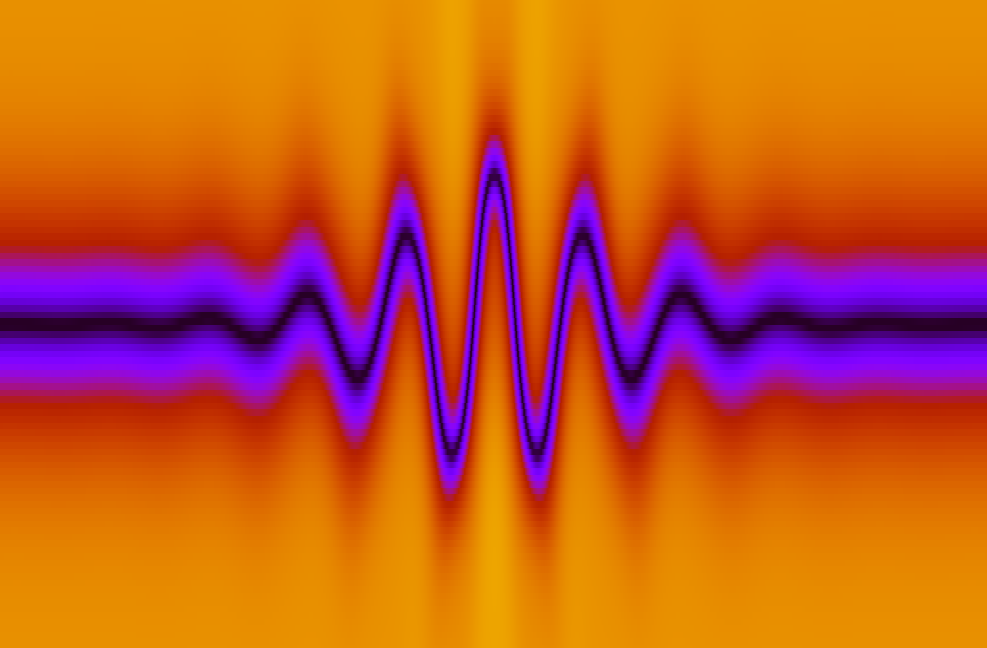}
\includegraphics[width=16cm, height=4cm]{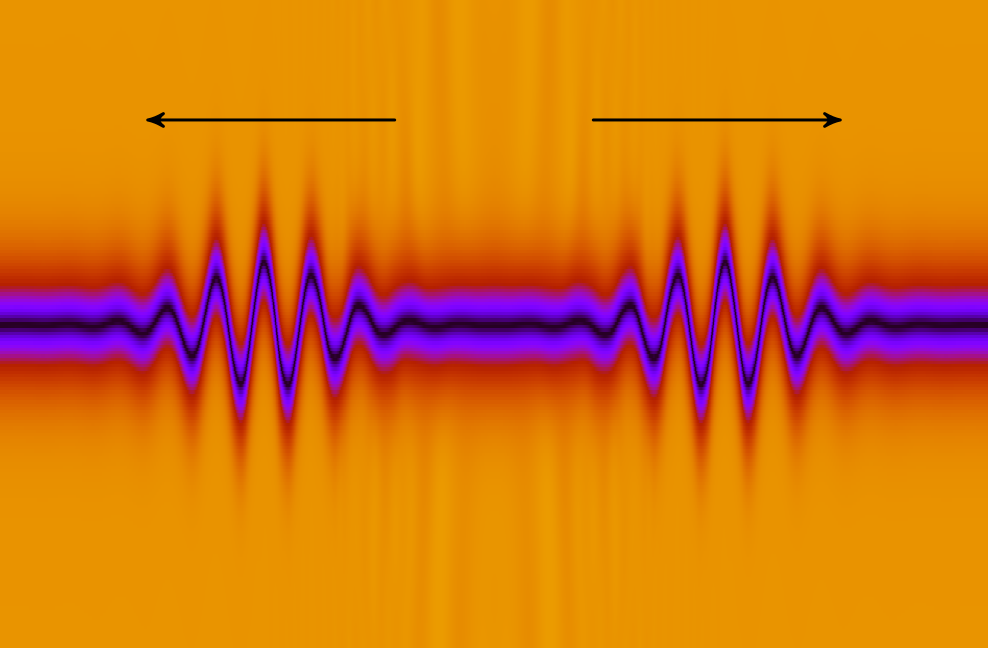}
\caption{Collision of two sinusoidal wave packets on the domain wall. We show the initial conditions
at the top panel, the moment of maximum overlap in the middle and a configuration after the
collision at the bottom. From top to bottom, the vertical axis $x$ ranges from $-8$ to $8$, 
$-4$ to $4$ and $-8$ to $8$, while the horizontal axis $y$ ranges from $-250$ to $250$, $-100$ to $100$ and $-200$ to $200$.}
\label{fig:collision cos wiggles}
\end{figure}

Let's now discuss what would be the evolution of these modes in a
cosmological setting. In \cite{Blanco-Pillado:2020smt} we showed numerically that cosmological phase transitions 
would  naturally lead to the formation of excited solitons. The typical amount of 
extra energy was found to be in the vicinity of $20\%$ of the relaxed soliton 
mass\footnote{Even though the simulations were performed in a simple $1+1$ dimensional lattice, we expect
a similar behaviour in a more realistic $3+1$ dimensional scenario to yield comparable results.}. It is
therefore clear that one would have a period of time where the dynamics of the solitons
could be affected by this extra energy by processes similar to the ones we discussed
in previous sections. However, this period of time is still very small compared
to the age of the universe, so we would like to ask whether there is any mechanism
that would allow for the cosmological network of defects to replenish the
level of excitation in the course of their evolution making these 
internal modes relevant for the long term dynamics of the real networks.

\begin{figure}[h!]
\includegraphics[width=13cm]{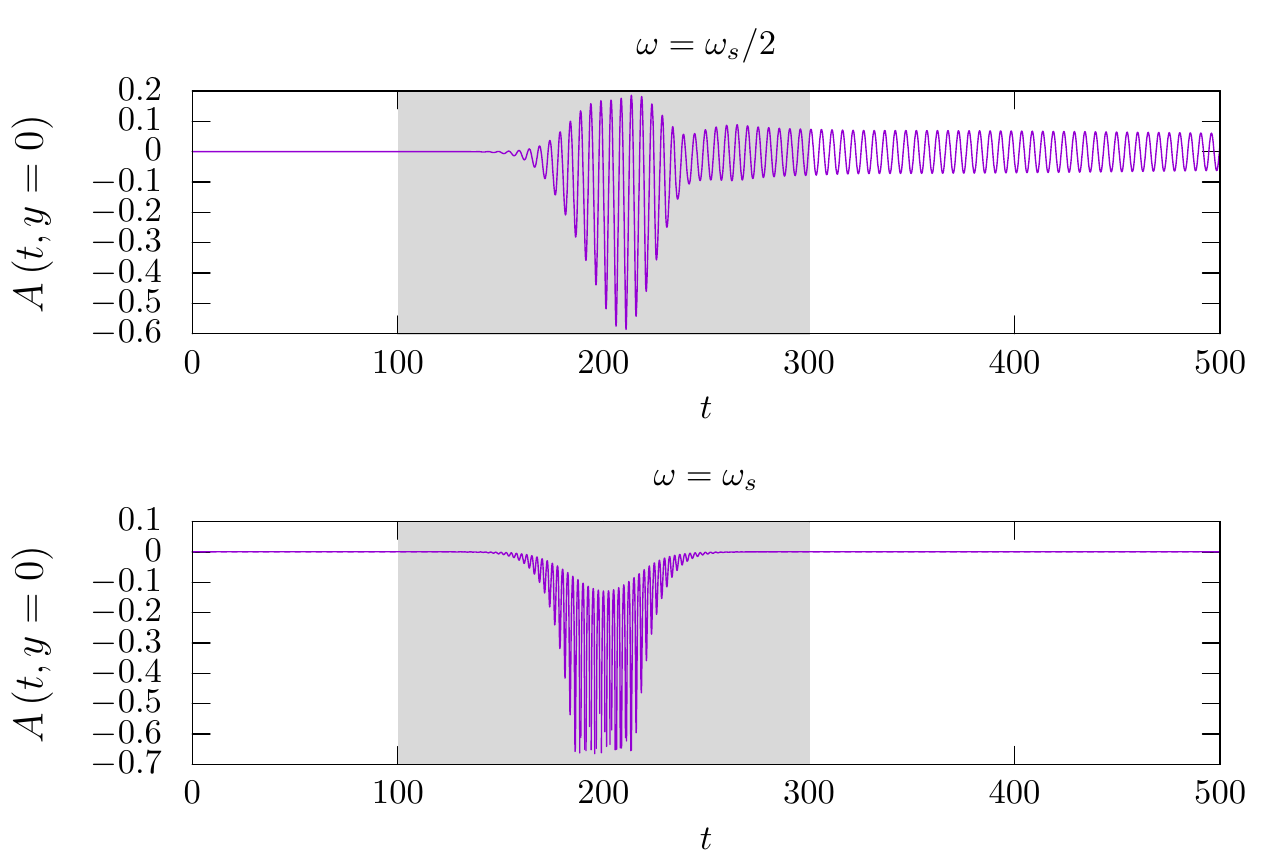}
\caption{Amplitude of the shape mode as a function of time at $y=0$, the center of the box. 
The shaded region represents the time during which the wiggles overlap. The internal mode remains 
excited after the collision only when the frequency of the wiggles is $\omega=\omega_{s}/2$.} 
\label{fig:excitation shape from collisions}
\end{figure}

In order to study this possibility, we first imagine the case where the 
domain wall is relaxed and ask the question of whether the interaction between
Goldstone modes could lead to a transfer of energy to the internal modes.
This is a similar situation to the one we explored before in order to check
the validity of the Nambu-Goto action, but now we are interested not only in the
dynamics of the colliding wiggles but on the domain wall state that they leave
behind. In particular, we will explore the collision between wiggles that are formed
by a localized wavepacket with a single frequency mode in the region of their
compact support. We take the initial conditions to be two of these packages 
separated by a distance in straight domain wall moving towards one another
at the speed of light. After the collision, we study the straight segment of the domain
wall left behind after the wiggles have passed through one another (see Fig. \ref{fig:collision cos wiggles}). One can analyze 
the amount of extra energy in this segment by projecting out the scalar field
configuration onto the shape mode. Doing this, one can compute the amount
of energy transferred from the wiggles to the internal modes as a function
of the typical frequency of the wave packets\footnote{Note that the
reconstruction of the excited mode during the time that the wiggles overlap (the shaded 
region in Fig. \ref{fig:excitation shape from collisions}) is problematic due to the rapid 
motion of the domain wall. However, we are only interested in the result after the 
collision, where this projection can be trusted. We have checked that our results do
not change by increasing the resolution in our lattice simulations.}.

Our results indicate that the internal mode is not excited by any appreciable amount
except in one particular situation. Taking the frequency of the wiggles to be
half of the frequency of the internal mode, one can find an excited segment of the
wall. This could be understood by looking at Eqs. (\ref{eq:reducedEqs1}) and (\ref{eq:reducedEqs2}). There we see that a zero mode
fluctuation acts as a driving source for the internal mode. When the two wiggles 
overlap, one forms a standing wave for a short period of time with the appropriate
frequency to resonantly excite the amplitude of the internal mode. Other frequencies
do not create an amplification of these modes, as shown in Fig. \ref{fig:excitation shape from collisions}.

This demonstrates that once the defects on the network have relaxed from their initial
excited state it would be difficult to dynamically transfer energy to the internal
modes since one would not expect a large amount of high frequency Goldstone
modes on the walls at late times.

\section{Conclusions and implications for cosmic strings}

We have studied the dynamics of domain wall solitons that appear in a scalar
field theory in $(2+1)d$. The lowest energy configurations of these solitons describe
extended objects invariant along one of the spatial dimensions. They appear
as line-like objects in our $(2+1)d$ setting, so we correspondingly name them domain wall
strings.

We start our discussion by identifying the low energy excitations of these objects. These
perturbations are divided in two different families of modes. The localized modes that represent
small excitations around the background soliton solution and the radiative modes that describe
vacuum excitations that propagate freely to infinity. Looking at the localized modes, we
also distinguish two different types of perturbations. The Goldstone modes that parametrize
the propagation of wiggles on the domain wall strings and the so-called shape mode that
represents a change in the soliton's width. 

We start by studying how the localized modes on the domain wall string lead to radiation.
We obtain an analytic expression for the power emitted by the presence of wiggles
on the string. In particular, we look at standing waves on the position of the string in the
small amplitude regime. In this case we note that there is no radiation for wiggle excitations
with frequency lower than $w_0 = m/2$, where $m$ is the mass of the perturbative excitations
of the field in the vacuum. For larger frequencies, the power increases but it is cut-off
at $w>>m$. This is due to the destructive interference of the different sections of the
string that act as a source of the radiation.

We proceed in a similar manner to compute the radiation from the shape mode excitations.
The results indicate that a shape mode on the domain wall string can live for a substantially 
long time. This could be much longer than the typical time scale of the soliton associated
with its width. These results suggest the possibility of obtaining an effective worldsheet
action that incorporates the dynamics of these modes in the regime where they do not
appreciably radiate.

We start this investigation by considering the effective action for the wiggles on the domain
wall string. In the absence of any massive mode we take the Nambu-Goto as the appropriate
action. In order to test this action we perform a series of numerical experiments where
we compare the results obtained in a lattice field theory evolution with the predictions 
from NG. Our results show that, in the regime where the radius of curvature of the
domain wall string is large compared to the thickness of the soliton, the position of the
string is described perfectly by the NG action.

We then move on to describe the coupling of the shape mode to the NG action. The simplest
way to do this is to imagine this excitation as a massive scalar field degree of freedom living
on the string worldsheet. Guided by these ideas, we write the action at the lowest
order that couples these modes. The presence of this new degree of freedom can affect
the dynamics of the domain wall string. We test this idea by looking at the motion of
a domain wall loop that naturally collapses under its own tension. We do this in two different
ways: we solve the field theory equations for a ring with cylindrical symmetry and 
compare the results to the ones predicted by our effective action. The results
in field theory show that the presence of an excitation can slow down the collapse of the domain
wall. This is, indeed, qualitatively captured by the effective action as well. In the
effective action one can see this effect as a mechanical backreaction of the
excitation on the domain wall. However, it is difficult to go beyond this qualitative
comparison due to the presence of radiation in the most interesting cases where
the shape mode is more relevant.

Another interesting aspect of this effective theory is the appearance of parametric 
resonant effects between the internal mode excitation and the wiggles of the domain wall string.
The physical origin of this effect is due to the oscillatory behaviour of the effective
pressure on the domain wall string due to the scalar field excitations. This is a
similar effect to the one in the non-relativistic string discussed long ago \cite{Rayleigh}. This seems
to suggest that this may be a fairly generic effect that could be relevant for
other areas of physics where an internal mode changes periodically the
equation of state of a soliton. Applications to other effective theories
similar to the one presented here could be interesting, for example in the
context of braneworlds.

Our numerical investigations of these field theory excited solitons in a lattice show that
indeed these parametric resonance effects are present in the full theory. However,
a detailed comparison of the results shows that the frequency of the
zero mode instability is half of the one predicted in the effective theory.

We have studied this issue from a different perspective with the help
of a reduced Lagrangian that describes the coupling between the amplitudes
of the different field theory modes on the domain wall strings. This investigation has led us
to a Mathieu type equation for the amplitude of a standing wave of the string
position. This equation is well known to have regions of instability. This is
precisely the instability that we observe in field theory with the correct
frequency. 

These results suggest that one should include new couplings into the simplest 
effective theory that we have been discussing. One such term could be
a non-minimal coupling between the massive scalar field and the Ricci
scalar of the worldsheet. Looking at the lowest order approximation of
the equations of motion obtained from an action including this term,
one identifies the existence of the parametric resonances of the form
encountered in field theory. This hints to the possibility that indeed this
term should be present in the effective theory.

The appearance of these parametric resonances leads to a more rapid decay
of the energy stored in the shape mode. This seems to indicate that these
excited states would only play a transient role in a cosmological model. However,
this conclusion may change if there is any mechanism that would populate
these modes later on in the evolution of the defects. In order to investigate this
possibility, we have performed a numerical experiment in field theory that mimics
the possible excitation of these modes in the late universe. The process
assumes the presence of wiggles propagating on the domain wall that,
through their collision, could excite the internal modes. This is certainly
a possibility. However, our field theory simulations demonstrate that
these internal modes are not excited unless the frequency of the wiggles
is of the order of the thickness of the string. In other words, long wavelength
excitations expected to be present at late times do not seem to have
the capacity of populating the internal modes, so we conclude again that
internal modes are only relevant for a short cosmological time scale \footnote{We 
should show this explicitly by performing numerical simulations of networks
of domain wall strings in field theory. Some effort in this direction is already
underway.}.

Finally, let us remark that even though our arguments are very suggestive
they are not straightforwardly extrapolated to $(3+1)$ dimensions. In the
case of strings in  $(3+1)$d there are other processes that might be
relevant for the possible excitation of the internal modes. In particular,
it was suggested in \cite{Saurabh:2020pqe} that intercommutation processes (when
a loop is formed) could lead to internal excitations. It is hard to see 
whether this can be enough to significantly affect the evolution of the
loop as a whole, but this is something that should be investigated 
further. Furthermore, even though it seems reasonable to expect that
there will also be parametric resonant effects in these models in 
higher dimensions, this has not been conclusively shown yet. One should
therefore use similar numerical techniques to the ones presented in 
this paper to analyze these possible resonances in models of strings
in $(3+1)d$. We leave these issues for a future publication.

\section{Acknowledgements}

We are grateful to Ken Olum, Tanmay Vachaspati and Alex Vilenkin for many useful discussions.  
This work is supported in part by the Spanish Ministry MCIU/AEI/FEDER 
grant PGC2018-094626-B-C21 as well as the PID2021-123703NB-C21 grant funded by MCIN/
AEI /10.13039/501100011033/ and by ERDF; ”A way of making Europe”, 
 the Basque Government grants (IT-979-16 and IT-1628-22) and the Basque Foundation for Science (IKERBASQUE). The 
numerical work carried out in this paper has been possible thanks to the computing infrastructure 
of the ARINA cluster at the University of the Basque Country, (UPV/EHU).

\appendix

\section{Details of the numerical calculations}
\label{appendix-numerics}

In the following appendix we will describe some of the technical details about the lattice field theory
simulations in several different sections of the main paper. Furthermore, we will also give a explicit 
account of the way one can extract the relevant data to make a precise comparison of the evolution of the
domain wall strings with the predictions from the Nambu-Goto action.

In the rest of the section, $x$ will denote the direction perpendicular to the domain wall while $y$ would 
describe the parallel one. Moreover, we will refer to the length of each side of the simulation box as 
$L_{x}$ and $L_{y}$ (where $x,y$ range from $-L_{x,y}/2$ to $L_{x,y}/2$), and $\Delta x=\Delta y$ 
and $\Delta t$ will respectively denote the lattice spacing and time step. The number of points in each 
dimension is $L_{x,y}/\Delta x$. The numerical values of these scales will be chosen conveniently 
depending on the particular experiment we want to perform.

\subsection{Lattice simulations}

As we describe in the main part of the text, the equation of motion that we are 
solving throughout this paper can be written in dimensionless form as
\begin{equation}
\frac{\partial^{2}\phi}{\partial t^2}-\frac{\partial^{2}\phi}{\partial x^2}-\frac{\partial^{2}\phi}{\partial y^2}+\phi\left(\phi^{2}-1\right)=0 ~,
\label{eq:eom}
\end{equation}
where $\phi(t,x,y)$ describes the scalar field living in a $(2+1)d$ spacetime. Regarding the boundary
conditions, we use absorbing boundary conditions in the $x$ direction (for the details, see the following subsection) 
and periodic ones in the $y$ direction. Standard discretization techniques were employed for the numerical 
evolution (staggered leapfrog method and nearest neighbours), and we implemented \emph{message passing interface} 
(MPI) in our parallelized code in order to handle the large number of lattice points, which is typically of the order of $10^{6}$.\\\\

Let us now briefly discuss some of the specific details relevant for the different simulations we carry 
out in the sections of the main paper. In what follows, $L_{x,y}$ are to be compared with the dimensionless width of the string, $\delta\sim1$.

\begin{enumerate}
\item

In section IIIA we analyze the radiation emitted from small amplitude standing waves in the position of the 
core of the domain wall. As one can show analytically, when these perturbations couple quadratically to the 
scattering states, the wavelength of the radiation in the $x$ direction is approximately given 
by $\lambda_{rad}=2\pi/\sqrt{4\omega^{2}-2}$, where $\omega$ is the angular frequency of the standing 
wave. For $\omega$ sufficiently close to threshold frequency, $m/2=\sqrt{2}/2$, this wavelength is large, so 
one needs a big box in the transverse direction in order for the emitted radiation to fit in. Moreover, we are 
interested in the radiation very far away from the soliton. In our simulations (red points in Fig. \ref{fig:power zero mode}) 
we chose $L_{x}=200$, which is more than enough to avoid any problems related to the spatial extent of the 
radiation. On the other hand, we varied $L_{y}$ from $4$ to $26$ in order to get the desired frequencies for the 
standing wave \footnote{Recall that periodic boundary conditions in the $y$ direction force waves on the string to 
have frequencies given by $2\pi n/L_{y}$, where $n$ is an integer number.}. Regarding lattice spacing, we 
used $\Delta x=0.02, 0.04$ and $\Delta t=0.005, 0.01$. We ran the simulations for $2\times10^{5}$ time steps 
and found the energy flux accross the lines $x=\pm98$ at late times. As we will point out in the following 
subsection, the refinement of the absorbing boundary conditions taking into account the particular frequency 
of the emitted radiation was crucial in these experiments. \\

\item For the simulations of travelling wave collisions in section IVA1, we need a large initial separation between wiggles 
in order for them to be as close as possible to exact Vachaspati-Vachaspati solutions  \cite{Vachaspati:1990sk}. We evolve 
the initial configuration Eq. (\ref{eq:psi plus psi minus}) with
\begin{equation}
\psi_{+}\left(y+t\right)=-8\left[\tanh\left(\frac{y+t-120}{12}\right)-\tanh\left(\frac{y+t-80}{12}\right)\right],
\label{eq:psi plus}
\end{equation} 
\begin{equation}
\psi_{-}\left(y-t\right)=8\left[\tanh\left(\frac{y-t+120}{12}\right)-\tanh\left(\frac{y-t+80}{12}\right)\right].
\label{eq:psi minus}
\end{equation} 
\\
The box size and lattice spacing in these simulations were $L_{x}=100$, $L_{y}=400$ and 
$\Delta x=0.1$; $\Delta t=0.05$.  We ran for thousands of time steps in order to let the wiggles 
collide several times.

These wiggles are quite flat even for big amplitudes, which is convenient because we want to 
avoid high curvatures that could spoil the comparison with the Nambu-Goto dynamics. For these 
particular wiggles, the curvature (i.e., the ratio string thickness to radius of curvature) during interaction 
is around $0.12$ (at most).\\

\item
In the simulations of section VI, the initially excited shape mode triggers the amplification of zero 
modes with specific frequencies. In order to see this effect, we had to make sure that the size of the
 box in the $y$ direction was a multiple of the wavelength of the resonant zero modes.
 
For the experiments in section VIB, we start out with the configuration given in Eq. (\ref{eq:dw plus shape 
 plus zero}) with $\omega_{0}=2\pi/10$ (which is sufficiently close to the actual resonance 
 frequency, $\omega_{s}/2$) and a very small amplitude for the zero mode in order to accelerate the
 apperance of the mode amplification. According to Eq. (\ref{eq:reducedEqs2}), if this amplitude is set initially to $0$, the 
 zero mode should not be amplified. However, numerically the resonance occurs, albeit at very 
 late times. Indeed, numerical inaccuracies are enough to excite the resonance. For these simulations 
 we chose $L_{x}=L_{y}=20$, $\Delta x=0.01$ and $\Delta t=0.005$.

In section VIC the field was initialized according to Eq. (\ref{eq:dw plus shape plus zero plus zero}). We  
performed two groups of simulations corresponding to two different wave numbers for the shape 
mode: $k_{s}=2\pi/3.4$ and $k_{s}=2\pi/5$. For each group we took several values for the length 
of the box in the $y$ direction: $L_{y}=17,34,51,68$ for the former and $L_{y}=20,50,75,85,100$ for 
the latter, with typical lattice spacing of $\Delta x=0.03$ and time step $\Delta t=0.01$.\\

\item
Finally, the collisions of travelling wave packets in section VII were simulated in a lattice with 
$L_{x}=32$, $L_{y}=800$, $\Delta x=0.08$ and $\Delta t=0.02$. The initial state is given by 
Eq. (\ref{eq:psi plus psi minus}) with
\begin{equation}
\psi_{+}\left(y+t\right)=-\frac{B}{4}\left[\tanh\left(\frac{y+t-220}{25}\right)-\tanh\left(\frac{y+t-180}{25}\right)\right]\cos\left[\omega\left(y+t\right)\right]\,,
\label{eq:psi plus cow}
\end{equation} 
\begin{equation}
\psi_{-}\left(y-t\right)=\frac{B}{4}\left[\tanh\left(\frac{y-t+220}{25}\right)-\tanh\left(\frac{y-t+180}{25}\right)\right]\cos\left[\omega\left(y-t\right)\right]\,.
\label{eq:psi minus cow}
\end{equation} 
\\
These are sinusoidal wiggles with amplitude controlled by $B$ and angular frequency $\omega$. We 
performed three groups of simulations: wiggles with frequency $\omega=2\pi/20$ (below the resonance frequency), 
$\omega=2\pi/10$ (very close to the resonance frequency) and $\omega=2\pi/5$ (above the resonance frequency). In 
each case, $B$ was adjusted to get the desired ratio of the thickness of the soliton to the radius of curvature of the order
of $ 0.1$.\\\\
We performed a convergence test to show the robustness of the results presented in that section, specifically in Fig. \ref{fig:excitation shape from collisions}. Simulations with $10^{6}$, $4\times 10^{6}$ and $1.6\times 10^{7}$ lattice points and $\Delta x=0.16,\,0.08\,\,\text{and}\,\,0.04$, respectively, give almost identical results.
\end{enumerate}
\subsection{Tuned Absorbing boundary conditions}

As previously stated, we have implemented absorbing boundary conditions in the $x$ direction. This is because
an important part of the radiation produced in our simulations is emitted along that direction. Implementing
these absorbing boundary conditions allows us to run for very long periods of time without having to worry 
about the effects of the radiation bouncing off the walls and re-exciting the soliton.

Let us briefly discuss the reasoning behind these specific boundary conditions we use. In order to do that,
we first note that far away from the soliton (in the $x\rightarrow\pm\infty$ limit) we can assume that the field 
is described as a perturbation around the vacuum of the form $\phi_{\pm}\left(x,y,t\right)=\pm1+\xi_{\pm}\left(x,y,t\right)$ . 
At the linear level, the equation of motion yields a solution for this perturbation as a travelling wave, 
$\xi_{\pm}\left(x,y,t\right)\propto\cos\left(\omega_{r}t\mp k_{x}x\mp k_{y}y+\delta\right)$ with 
$\omega_{r}=\sqrt{k_{x}^{2}+k_{y}^{2}+2}$ (recall that the dimensionless mass is $\sqrt{2}$). Assuming 
normal incidence (i.e., $k_{y}=0$), this wave is a solution to the so-called Mur boundary conditions \cite{Mur}, 
\begin{equation}
\frac{\partial\phi}{\partial t}\pm\frac{\partial\phi}{\partial x}\,\,\,\,\bigg\rvert_{x=\pm L_{x}/2}=0\,,
\label{eq:Mur abc}
\end{equation}
if $\omega_{r}=k_{x}$. Therefore, these conditions are efficient at absorbing waves with $k_{x}>>m$.

One can refine the Mur boundary conditions for monochromatic radiation with known angular frequency $\omega_{r}$
by using instead
\begin{equation}
\frac{\partial\phi}{\partial t}\pm\frac{\omega_{r}}{\sqrt{\omega_{r}^{2}-m^{2}}}\frac{\partial\phi}{\partial x}\,\,\,\,\bigg\rvert_{x=\pm L_{x}/2}=0\,.
\label{eq:refined Mur abc}
\end{equation}
\\
As one can easily check, the travelling waves $\xi_{\pm}$ above are exact solutions to these equations. This tuning 
of the absorbing boundary conditions turned out to be crucial in our simulations, specially in those of section III, where 
we knew the angular frequency of the outgoing radiation: $\omega_{r}=2\omega$. Here, $\omega$ is the frequency 
of the zero mode that couples to the scattering states. Note that  (\ref{eq:Mur abc}) and (\ref{eq:refined Mur abc})  
are nearly equivalent if $\omega_{r}>>m$, or $\omega>>m/2$. However, this may not be the case as we are free 
to choose the angular frequency of the zero mode to be arbitrarily close to $m/2$. We employed the refined boundary 
conditions (\ref{eq:refined Mur abc}) for the standing wave experiments in sections III and IV, and the version (\ref{eq:Mur abc}) for the rest.

\subsection{Nambu-Goto reconstruction of a field theory domain wall}
\label{appendix-NGreconstruction}

In this appendix we would like to outline the main steps required to obtain a Nambu-Goto
description of a domain wall string that has been simulated in a lattice field theory \footnote{Here we use the term
Nambu-Goto to mean that we can parametrize the domain wall string with the usual conformal gauge in 
a Nambu-Goto description (see section \ref{NG-section} for more details).}. The starting
point for this reconstruction is to obtain the data points for the position of the domain wall for
the particular time step that we are interested in. We do this in our lattice by identifying the spatial
coordinates on the ($x,y$) plane where the field goes to zero. This is our definition of the 
position of the domain wall.

In order to subsequently obtain the velocity of the domain wall string we actually need two data lists containing 
the spatial coordinates of the domain wall at $t=0$ and $t=\Delta t$:
\begin{itemize}
\item \emph{String 1}: list of $N$ vectors $\vec{r}_{n}^{\,\,0}$, with $n=0,1,...,N-1$, corresponding to the positions of every point on the domain wall string at $t=0$.
\item \emph{String 2}: list of $M$ vectors $\tilde{\vec{r}}_{m}^{\,\,0}$, with $m=0,1,...,M-1$, corresponding to the positions of every point on the domain wall string at $t=\Delta t$.
\end{itemize}
The Nambu-Goto reconstruction algorithm consists of the following steps:

\subsubsection{Computing tangent vectors}
The first step is to find the $N$ vectors $\vec{r}_{n}^{\,\,t}$ which are tangent to \emph{String 1} at each point $n$. For $n=0$ to $n=N-1$, we define the tangent vectors 
\begin{equation}
\vec{r}_{n}^{\,\,t}=\vec{r}_{n+1}-\vec{r}_{n}\,,
\label{eq:tangent vectors}
\end{equation}
where $\vec{r}_{N}=\vec{r}_{0}+L_{y}\hat{j}$. Here, $\hat{j}$ denotes the unitary vector in the $y$ direction. We will also need the unit tangent vectors, namely
\begin{equation}
\hat{\vec{r}}_{n}^{\,\,t}=\frac{\vec{r}_{n+1}-\vec{r}_{n}}{|\vec{r}_{n+1}-\vec{r}_{n}|} \,.
\label{eq:unit tangent vectors}
\end{equation}

\subsubsection{Nambu-Goto velocity and Lorentz factor}

Consider a particular point $n$ on \emph{String 1}. Let us call this point $P$. In order to find its velocity, we 
have to identify it with a point $P_{*}$ on \emph{String 2} (note that, in general, the point $P_{*}$ we want 
to find is not exactly one of the $M$ vectors found in the first step of the algorithm). Since in the conformal gauge in NG
the velocity of $P$ is perpendicular to $\vec{r}_{n}^{\,\,t}$, $P_{*}$ is the intersection of \emph{String 2} with the plane 
$\Pi_{n}$ perpendicular to $\vec{r}_{n}^{\,\,t}$ at $P$. Let $q$ and $q+1$ be the labels of the two points 
on \emph{String 2} (with position vectors $\tilde{\vec{r}}_{q}$ and $\tilde{\vec{r}}_{q+1}$) which are closer 
to $\Pi_{n}$. If they are in opposite sides of the plane, $P_{*}$ is the intersection of $\Pi_{n}$ with the line 
that contains both points $q$ and $q+1$. 

Once we have found the point $P_{*}$ on \emph{String 2} which corresponds to $\vec{r}_{n}$ on
 \emph{String 1}, we can compute the Nambu-Goto velocity as
\begin{equation}
\dot{\vec{r}}_{n}^{\,\,NG}=\frac{\tilde{\vec{r}}_{*}-\vec{r}_{n}}{\Delta t}\,\,,
\label{eq:NG velocity}
\end{equation}
and the Lorentz factor as
\begin{equation}
\Gamma_{n}=\frac{1}{\sqrt{1-|\dot{\vec{r}}_{n}^{\,\,NG}|^{2}}}\,\,.
\label{eq:lorentz factor}
\end{equation}

\subsubsection{Amount of $\sigma$ parameter in each string segment}
The gauge we are working in imposes the following condition:
\begin{equation}
\dot{\vec{r}}_{n}^{\,\,NG\,\,2}+\vec{r}_{n}^{\,\,'\,2}=1\,,
\label{eq:condition2}
\end{equation}
where the prime denotes derivative with respect to $\sigma$ (see beginning of section IV). 
Therefore, $|\vec{r}_{n}^{\,\,'}|=|\vec{r}_{n}^{\,\,t}|/\Delta\sigma_{n}=\sqrt{1-\dot{\vec{r}}_{n}^{\,\,NG\,\,2}}$, which yields
\begin{equation}
\Delta\sigma_{n}=\Gamma_{n}|\vec{r}_{n}^{\,\,t}|\,,
\label{eq:amount of sigma}
\end{equation} 
which gives the amount of energy (or amount of parameter $\sigma$) in this
section of the string.

\subsubsection{Redefinition of the tangent vectors}
Now, we simply redefine the tangent vectors according to their correct parametrization
in our gauge in the following way:
\begin{equation}
\vec{r}_{n}^{\,\,t\,NG}=\frac{\hat{\vec{r}}_{n}^{\,\,t}}{\Gamma_{n}}\,\,.
\label{eq:new tangent vectors}
\end{equation}

\subsubsection{Left and right movers: $\vec{a}\left(\sigma\right)$ and $\vec{b}\left(\sigma\right)$}
With the Nambu-Goto velocities (\ref{eq:NG velocity}) and the new Nambu-Goto tangent vectors (\ref{eq:new tangent vectors}), we can find the vectors
\begin{equation}
\begin{cases}
\vec{a}_{n}^{\,\,'}=\vec{r}_{n}^{\,\,t\,NG}-\dot{\vec{r}}_{n}^{\,\,NG}\, ,\\
\vec{b}_{n}^{\,\,'}=\vec{r}_{n}^{\,\,t\,NG}+\dot{\vec{r}}_{n}^{\,\,NG} ~.
\end{cases}
\label{eq:aprime and bprime}
\end{equation}
\\
Moreover, as we also know the amount of sigma parameter $\Delta\sigma_{n}$ in each string segment, we can obtain the correct
parametrization of $\vec{a}\left(\sigma\right)$ and $\vec{b}\left(\sigma\right)$. This is, in fact,
all the information we need to reconstruct the evolution of the string at any moment in time. We have
used this algorithm to obtain the motion of the domain wall string in section \ref{NG-section}.

Finally, let us comment on the reconstructed NG energy. At each time step, the energy of the string can be computed as 
\begin{equation}
E=\sum_{n=0}^{N-1}\Delta\sigma_{n}=\sum_{n=0}^{N-1}\Gamma_{n}|\vec{r}_{n}^{\,\,t}|\,,
\label{eq:energy}
\end{equation}
where we have set the string tension $\mu=1$. If the string moves according to the NG action, $E$ is conserved. 
Let us label our lattice field theory files as $1,2,...,T$, where $T$ is the total number of time steps. One can calculate 
the quantity (\ref{eq:energy}) taking these files in pairs ($\Gamma_{n}$ contains the velocities of each point, and the 
computation of the velocities requires two time steps) to obtain the NG energy at every time step in the field theory 
simulation: $E_{1}, E_{2},..., E_{T-1}$. Here, $E_{j}$ is the energy of the string computed from files $j$ and $j+1$. If these 
energies are different from each other, the string is not behaving as the NG action predicts.\color{black}

\section{The scalar field ansatz for Nambu-Goto dynamics}
\label{appendix-ansatz for NG dynamics}

In many instances, one would like to have an expression for the 
scalar field in our model that accurately represents 
a domain wall moving according to the Nambu-Goto dynamics.
Here, we use the knowledge of several examples where these
solutions are known exactly to motivate a scalar field ansatz
that approximately parametrizes the scalar field theory configuration from a generic 
solution of the Nambu-Goto dynamics.

Let's start by describing the static field theory configuration of the domain
wall centered around the $x=x_0$ point and extended in the $y$ direction, namely the solution
\beq
\phi_s(x,y,t) = \phi_K(x-x_0) ~.
\eeq

It is well known that one can obtain a new exact solution
describing a domain wall moving at a constant velocity ($v$) by
writing
\beq
\label{boosted}
\phi_b(x,y,t) = \phi_K\left(\frac{x-(x_0 + v t)}{\sqrt{1-v^2}}\right) ~.
\eeq
This is obviously an exact solution since it can be obtained by
boosting the original solution to a frame moving with velocity $v$.

Let us now consider another type of solution where the domain wall 
is not completely aligned with the $y$ axis but is slightly inclined with respect
to it. The solution for the static configuration of this kind would be given by
\beq
\label{rotated}
\phi_{\Theta}(x,y,t) = \phi_K\left( (x-x_0) \cos \Theta  -  y \sin \Theta  \right) =  \phi_K\left( \frac{x  - (x_0 + y \tan \Theta)}{\sqrt{1+\tan^2 \Theta}} \right)\,.
\eeq
This is nothing but the same static configuration obtained earlier but
now rotated by an angle $\Theta$. In the last expression the argument of the field $\phi_K$ has been expressed in a particular
way for reasons that will become apparent immediately.

Both these types of solutions given in Eqs (\ref{boosted}) and (\ref{rotated}) 
are exact solutions of the equations
of motion and can be collectively described by a single expression
of the form
\beq
\phi(x,y,t) = \phi_K\left( \frac{x- \psi(y,t)} {\sqrt{1 + \psi'^2 - \dot \psi^2}}\right)\,,
\eeq
if we take $\psi_b(y,t) = x_0 + v t$ for the boosted domain wall case and 
$\psi_\Theta(y,t) = x_0 + y \tan \Theta $ for the wall with an angle.

Moreover, we also notice that taking this ansatz for the scalar field with 
the extra condition that  $\psi'^2 = \dot \psi^2$ one identifies the 
exact expression for the Vachaspati-Vachaspati travelling wave 
solutions described in \cite{Vachaspati:1990sk}.

All these arguments strongly suggest that this type of ansatz could be a
good description for a generic motion of the string. Therefore, we 
consider the ansatz
\beq
\phi(x,y,t) = \phi_K\left( \frac{x- \psi(y,t)}{\sqrt{1- \partial_a \psi \partial^a \psi}}\right)\,,
\eeq
where we have introduced the relativistic notation for the coordinates
along the string worldsheet, namely $a=t, y$.

Guided by these considerations, we decided to study the effective
action for the field $\psi(y,t)$ in our model by plugging this
expression into the original $3d$ scalar field action. Doing this exercise, one
arrives to an action of the form
\beq
 S_{\text{effective}} = - \mu \int{dt ~dy \sqrt{1- \partial_a \psi \partial^a \psi}} + ...\,,
 \eeq
where the extra terms are propotional to higher order derivatives of the
field $\psi(y,t)$. This is exactly the Nambu-Goto action for the dynamics
of the perturbations transverse to a domain wall extended along the $y$
direction.

This demonstrates that our field theory ansatz should be an accurate
description for the full dynamics of the model as long as the field
$\psi(y,t)$ obeys the Nambu-Goto dynamics and one can neglect 
the higher order terms (curvature terms) and of course any source of 
radiation.

Finally, note that using this ansatz for the kink solution in $1+1$ dimensions 
one recovers the effective action for a relativistic particle dynamics. This
could be interesting in models that try to describe the motion of the
kink with the use of the collective coordinate language.

\section{Computing the radiation from domain wall excitations}
\label{appendix-radiation}

\subsection{Analytic calculation of the radiation from zero mode excitations}
\label{an_zero}

Let us consider the following scalar field configuration:
\be\label{ansatzJJ-shape}
\phi(t,x,y)= \phi_K \left[\frac{x-\psi_0(t,y)}{ \sqrt{1-\pa_\mu\psi_0 \pa^\mu\psi_0}} \right]+R(t,x,y)\,,
\ee
where $\phi_K$ is the static domain wall solution, $R(t,x,y)$ denotes collectivelly the radiation modes and $\psi_0(t,y)$ denotes a zero mode of frequency $\omega_0$ whose expression is given by
\be
\psi_{0}(t,y)=\hat{D} (t) \cos(\omega_0 t)\cos(\omega_0 y)~.
\ee
If we substitute (\ref{ansatzJJ-shape}) into the field equation (\ref{eq:eom}) we obtain, at second order in the amplitude $\hat{D}(t)$, the following equation for the radiation modes:
\be\label{eq:radiation}
\square R+\left(3\phi_K^2-1\right)R=-\frac{1}{\sqrt{2}}
\hat{D}^2 \omega_0^4 x\,
   \text{sech}^2\left(\frac{x}{\sqrt{2}}
   \right) (\cos (2 \omega_0 t)+\cos (2 \omega_0
   y))\,.
\ee
In order to obtain the expression (\ref{eq:radiation}) we have kept only linear terms in $R(t,x,y)$, since the radiation has a quadratic source in $\hat{D}(t)$. The response of $R(t,x,y)$ to the time independent term in the inhomogeneous part of  (\ref{eq:radiation}) will be itself time independent and it will carry no energy. We may therefore consider the time dependent part as source term for radiation and apply the Green's function method \cite{Manton:1996ex,Blanco-Pillado:2020smt} to obtain an asymptotic expression for $R(t,x,y)$ whose form is
\be
R=\frac{\hat{D}^2\pi   \omega_0^4 q
   \text{csch}\left(\frac{\pi 
   q}{2}\right)}
   {2 \sqrt{q^4+5
   q^2+4}} \cos \left(
  \omega t-\frac{q x}{\sqrt{2}} 
     - \tan ^{-1}\left(\frac{3
   q}{q^2-2}\right)\right)\,,
\ee
where 
\be
\omega=2   \omega_0,\,\quad q=\sqrt{2  \omega^2-4}\,.
\ee

The averaged radiated power per unit length at infinity in the $x$-direction can be obtained from the energy-momentum tensor:
\be\label{eq:average power per unit length}
\frac{\langle\dot{E}\rangle}{L_{y}}=\frac{\langle T_{0x}\rangle}{L_{y}}=-\frac{\hat{D}^4 \pi^2 \omega_0^9 q^3  \text{csch}^2\left(\frac{\pi q}{2}\right) }{4 \sqrt{2}\left(q^4+5 q^2+4\right)}\,.
\ee
where $L_y$ denotes the length of the domain wall string along the extended direction.

On the other hand, the total energy at quadratic order in $\hat{D}$ for a field configuration of the form (\ref{ansatzJJ-shape}) is the following:
\be
E(t)=M_{DW}+\frac{\sqrt{2}}{6}\omega_{0}^2 L_{y} \hat{D}^2(t)+\mathcal{O}(\hat{D}^3(t))\,,
\ee
where $M_{DW}=\sqrt{\frac{8}{9}}L_{y}$ is the mass of the static domain wall solution. If we assume that all the energy radiated to infinity comes from the zero mode, we obtain the following relation:
\be\label{Eq:EA}
\frac{dE(t)}{dt}=\langle\dot{E}\rangle\,,
\ee
and therefore,
\be
\frac{\sqrt{2}}{3}\omega_0^2 \frac{d}{dt} \hat{D}^2(t)=-\frac{ \pi^2 \omega_0^9 q^3  \text{csch}^2\left(\frac{\pi q}{2}\right) }{4 \sqrt{2}\left(q^4+5 q^2+4\right)}\hat{D}^4(t)\equiv-\alpha(\omega_0)\hat{D}^4(t)\,,
\ee
which can be integrated to give
\be
\hat{D}(t)=\frac{1}{\sqrt{\frac{1}{\hat{D}^2(0)}+\frac{3\alpha(\omega_0)}{\sqrt{2}\omega_0^2}t}}\,.
\ee


\subsection{Analytic calculation of the radiation from the internal mode excitations}
\label{an_shape}

We follow a similar procedure to that of Appendix \ref{an_zero} to obtain the radiation emitted by the domain wall string shape modes. Let us consider the following ansatz:
\be\label{ansatzJJ-shape-1}
\phi(t,x,y)= \phi_K \left(x \right)+\hat{A}\cos\left(\omega t\right) f_1(x)\cos(ky)+R(t,x,y)\,,
\ee
where $R(t,x,y)$ corresponds to the radiation modes, $f_1(x)$ is the shape mode profile and $k$ is the frequency of the mode in the longitudinal direction. At second order in $\hat{A}$ we obtain the following equation for the radiation:
\bea
&&\square R+\left(3\phi_K^2-1\right)R=-3\phi_K \hat{A}^2f_1^2\cos^2(\omega t)\cos^2(k y)=\nonumber\\
&&-\frac{3}{4}\phi_K \hat{A}^2f_1^2\cos(2\omega t)\cos(2k y)-\frac{3}{4}\phi_K \hat{A}^2f_1^2\cos(2\omega t)+\text{(time independent terms)} \,,~~~~\label{rad11}
\eea
where $\omega=\sqrt{\omega_s^2+k^2}$ and $\omega_s=\sqrt{3/2}$. As we have mentioned, the response of the radiation to time independent sources is itself time independent and therefore does not carry away radiation. The time dependent term suggests the following form of radiation at infinity:
\be
R(t,x,y)=R_1(x)\cos(2\omega t)+R_2(x)\cos(2\omega t)\cos(2 k y)\,.\label{ansatz11}
\ee
Now,  (\ref{rad11}) is satisfied for the ansatz (\ref{ansatz11}) if the following equations are satisfied simultaneously:
\bea
-R_1''(x)+\left(3\phi_K^2-1-(2\omega)^2\right)R_1&=&-\frac{3}{4}\phi_K \hat{A}^2f_1^2\,,\\
-R_2''(x)+\left(3\phi_K^2-1-(2\omega_s)^2\right)R_2&=&-\frac{3}{4}\phi_K \hat{A}^2f_1^2\,.
\eea
The radiation at infinity may be obtained with the Green's function method \cite{Manton:1996ex,Blanco-Pillado:2020smt}. For $R_2$ we get
\be
R_2(t,x,y)=\frac{3}{16} \sqrt{3} \pi  \hat{A}^2
   \text{csch}\left(\sqrt{2} \pi \right) \cos
   \left(\sqrt{6} t-2 x-\tan
   ^{-1}\left(\sqrt{2}\right)\right)\cos(2 k y)\,.
\ee
For $R_1$ we get
\be
R_1(t,x,y)=\frac{3 \pi  \hat{A}^2 q \left(q^2-2\right)
   \left(q^2+4\right) \text{csch}\left(\frac{\pi 
   q}{2}\right) \cos \left(-\tan ^{-1}\left(\frac{3
   q}{2-q^2}\right)+\frac{q x}{\sqrt{2}}- t \bar{\omega}
   \right)}{128 \sqrt{2} \sqrt{q^4+5 q^2+4}}	\,,
\ee
where $q=\sqrt{2\bar{\omega}^2-4}$ and $\bar{\omega}=2\omega$.  The averaged radiated power per unit length at infinity in the $x$-direction can be obtained again from the energy-momentum tensor,
\be\label{EM:shape}
\langle T_{0x}\rangle=\langle\left(\pa_t R_1+\pa_t R_2\right)\left(\pa_x R_1+\pa_x R_2\right)\rangle_{t,y}\,.
\ee
Note that for $k=0$ we recover the standard expression of the averaged radiated power per unit length at infinity $\langle T_{0x}\rangle=\langle T^{\text{(kink)}}_{0x}\rangle $, (see \cite{Manton:1996ex,Blanco-Pillado:2020smt}). For $k\neq 0$ the cross terms in (\ref{EM:shape}) are eliminated by the average in the y-direction, reducing the averaged radiated power to the following expression:
\be
\langle T_{0x}\rangle=\langle \pa_t R_1\pa_x R_1+\pa_t R_2\pa_x R_2\rangle_t \equiv T_1+T_2\,.
\ee
The expression for $T_2$ does not depend on the frequency in the longitudinal direction and can be related to the averaged power of the standard kink as follows:
\be
T_2=\frac{1}{8}T_{0x}^\text{(kink)}=-\frac{27}{128} \sqrt{\frac{3}{2}} \pi ^2\hat{A}^4
   \text{csch}^2\left(\sqrt{2} \pi \right)=-0.0014 \hat{A}^4\,.
\ee
On the other hand, $T_1$ does depend on the longitudinal frequency.  A similar calculation shows that
\be
T_1=-\frac{9 \pi ^2 \hat{A}^4 q^3 \left(q^2-2\right)^2
   \left(q^2+4\right)^3
   \text{csch}^2\left(\frac{\pi  q}{2}\right)}{65536
   \sqrt{2} \left(q^4+5 q^2+4\right)}\,.
   \ee
 Finally, taking into account (\ref{Eq:EA}), we can relate the rate of the energy loss of the shape modes with the radiated power at infinity: 
 \be
 \frac{1}{2}\left(\frac{3}{2}+k^2\right)\frac{d}{dt}\hat{A}^2(t)=(T_1+T_2)\hat{A}^4(t)\,,
 \ee
which after integration gives
\be\label{eq:manton law}
\hat{A}(t)=\frac{1}{\sqrt{\frac{1}{\hat{A}^2(0)}+\frac{2(T_1+T_2)}{\omega_s^2}t}}\,.
\ee


\section{The Lagrangian for the interacting mode amplitudes}
\label{appendix-lagrangians}

In this section we will study the interaction between zero modes and internal modes. The general domain wall solution 
with periodic boundary conditions in the $y$-direction in the box $\mathbb{R}\times [-L_{y}/2,L_{y}/2]$ can be expanded 
in the basis of linearized modes as follows:
\bea\label{gen:sol}
\phi(t,x,y)&=&\phi_K(x)+\sum_{k=0}^\infty D_k(t)f_0 (x)\cos (\omega_0(k) y)+\\
&+&\sum_{k=0}^\infty A_k(t)f_1 (x)\cos (\omega(k) y)+R(t,x,y)\,,\nonumber
\eea
where
\be
\omega_0(k)=\frac{2\pi k}{L_{y}},~\omega(k)=\sqrt{\omega_s^2+\left(\frac{2\pi k}{L_{y}}\right)^2}~,
\ee
and $R(t,x,y)$ stands for the radiation modes. The ansatz (\ref{gen:sol}) contains all the degrees of freedom and as a consequence, is a general solution of the field equations in the frame of the domain wall. 

Let us assume now that the relevant degrees of freedom of the system are given by the zero mode and the homogeneous shape mode (i.e. the shape mode with $k=0$). We consider the following ansatz:
\be\label{ansatz_2}
\phi(t,x,y)=\phi_K(x)+A(t)f_1 (x)+D(t)f_0 (x)\cos (\omega_0 y)	\,.
\ee
This choice of ansatz implicitly assumes that the radiation does not play any relevant role during the mode interaction. After a proper rescaling of the domain wall action we may substitute (\ref{ansatz_2}) into (\ref{Act:DW}). Integration in the $x,y$-directions gives the following effective mechanical Lagrangian with two degrees of freedom:
\bea
\frac{2}{L_{y}}\mathcal{L}_{\text{effec}}&=&-\frac{4\sqrt{2}}{3}+\dot{A}^2(t)-\frac{3}{2}A^2(t)+\frac{1}{2}\dot{D}^2(t)-\frac{1}{2}\omega_0^2 D^2(t)-\frac{3}{2}C_1 A^2(t)D^2(t)\\ \nonumber
&&-\frac{3}{4}C_1 A^4(t)-\frac{9}{16}C_1 D^4(t)-3C_2 A(t)D^2(t)-4 C_2A^3(t)\,,
\eea
where $C_1=3\sqrt{2}/35$ and $C_2=3\sqrt{3}\pi/(64 \,2^{3/4})$. The equations of motion for A(t) and D(t) are therefore
\beq
\label{eq:reducedEqs1:ap}
\ddot A(t) +\frac{3}{2} \left[ 1+ C_1 D^2(t) \right] A(t)+ 6 C_2 A^2(t) + \frac{3 C_1}{2} A^3(t) +  \frac{3 C_2}{2} D^2(t) = 0\,,
\eeq
\beq
\label{eq:reducedEqs2:ap}
\ddot D(t) + \left[ \omega_{0}^2 + 6 C_2 A(t)+ 3C_1 A^2(t) \right] D(t)  + \frac{9 C_1}{4} D^3(t)  = 0\,.
\eeq

Let us assume first that the shape mode is initially excited. Af first order we have
\be
A(t)=\hat{A}\cos(\omega_s t)\,.
\ee
In the background of this solution and assuming that $A(t)$ and $D(t)$ are small, the field equation for $D(t)$ takes approximately the following form:
\be\label{Eq:Mat}
\ddot{D}(t)+\left(\omega_0^2+6C_2 \hat{A}\cos(\omega_st)\right)D(t)= 0\,.
\ee

This is nothing but a Mathieu equation (see Appendix \ref{matsec}). It is well-known that this equation has instabilities  for some relations between  $\omega_0$, $\omega_s$ and $\hat{A}$. As we will explain later, in our case this translates into the condition $\omega_0\approx \frac{\omega_s}{2}$.  At this point, one expects an exponential growth in the amplitude of the zero mode.

On the other hand, if one assumes that the zero mode is initially excited,
\be
D(t)=\hat{D} \cos(\omega_0 t)\,,
\ee
then for initially small $D(t)$ and $A(t)$ the equation for the shape mode is
 \be
 \ddot A(t) +\frac{3}{2}  A(t)+  \frac{3 C_2}{2} \hat{D}^2\cos^2(\omega_0 t) =0\,.
 \ee
 This is the equation of the forced harmonic oscillator whose resonance will occur at $\omega_0=\omega_s/2=\sqrt{3/2}/2$. In a realistic situation both modes are coupled, allowing for an energy transfer between them. As we have explained is Sec. \ref{subsec:nonlinear}, this coupling prevents the amplitudes $A(t)$ or $D(t)$ to grow indefinitely.\\\\
 
 Let us take now the following initial configuration:
 \be
 \phi(t,x,y)=\phi_K(x)+A(t)f_1(x)\cos(k_s y)+D_1(t)f_0(x)\cos(k_1 y)+D_2(t)f_0(x)\cos(k_2 y)\,,
 \ee 
 i.e. the shape mode may have now a frequency $k_s$ in the $y$-direction, and two zero modes with different frequencies $k_1$ and $k_2$ are allowed. Repeating the same process as before we obtain the following mechanical Lagrangian:
\bea
\frac{2}{L_{y}}\mathcal{L}_{\text{effec}}&=&-\frac{4\sqrt{2}}{3}+\frac{1}{2}\dot{A}^2(t)+\frac{1}{2}\dot{D}_1^2(t)+\frac{1}{2}\dot{D}_2^2(t)-\frac{1}{2}(\omega_s^2 +k_s^2)A^2(t)-\\ \nonumber
&&-\frac{1}{2}k_1^2 D_1^2(t)-\frac{1}{2}k_2^2 D_2^2(t)-C_3\left(\frac{1}{2}A^4(t)+D_1^4(t)+D_2^4(t)\right)-\\ \nonumber
&&-\frac{3}{4}C_1\left(A^2(t)D_1^2(t)+A^2(t)D_2^2(t)+3D_1^2(t)D_2^2(t)\right)-3 C_2 A(t)D_1(t)D_2(t)\,,
\eea
where $C_3=27/(280\,\sqrt{2})$. The field equations at quadratic order are
\bea
\ddot A(t) + \left[ \omega_s^2+ k_s^2 \right] A(t)+  3C_2 D_1(t) D_2(t) &=& 0~,\\
\ddot D_1(t) +  k_1^2 D_1(t)  + 3C_2 A(t) D_2(t)  &=& 0~,\\
\ddot D_2(t) + k_2^2  D_2(t)  + 3C_2 A(t) D_1(t)  &=& 0~.
\eea

We have obtained the Lagrangian and field equations assuming that $k_1\neq k_s/2$, $k_2\neq k_s/2$ and $\vert k_1-k_2 \vert =k_s$. When the three conditions are met, it is possible for a shape mode with frequency $k_s$  in the $y$-direction to excite a couple of zero modes of frequencies $k_1$ and $k_2$ if, in addition, the condition $k_1+k_2=\sqrt{3/2+k_s^2}$ holds. For details in the study of these types of instabilities we refer to \cite{METTLER1967169}.  We show in the main part of the
text how one can find examples of these parametric resonances in multiple
modes in our field theory simulations of the domain wall string. The results
are in perfect agreement with the conclusions obtained from these equations.



\section{Brief review of the Mathieu equation}
\label{matsec}

The standard form of the Mathieu equation is given by
\be\label{Mathieu}
\ddot{x}(t)+\left(a-2q  \cos(2 t)\right)x(t)=0\,.
\ee

\begin{figure}[h!]
\includegraphics[width=10cm]{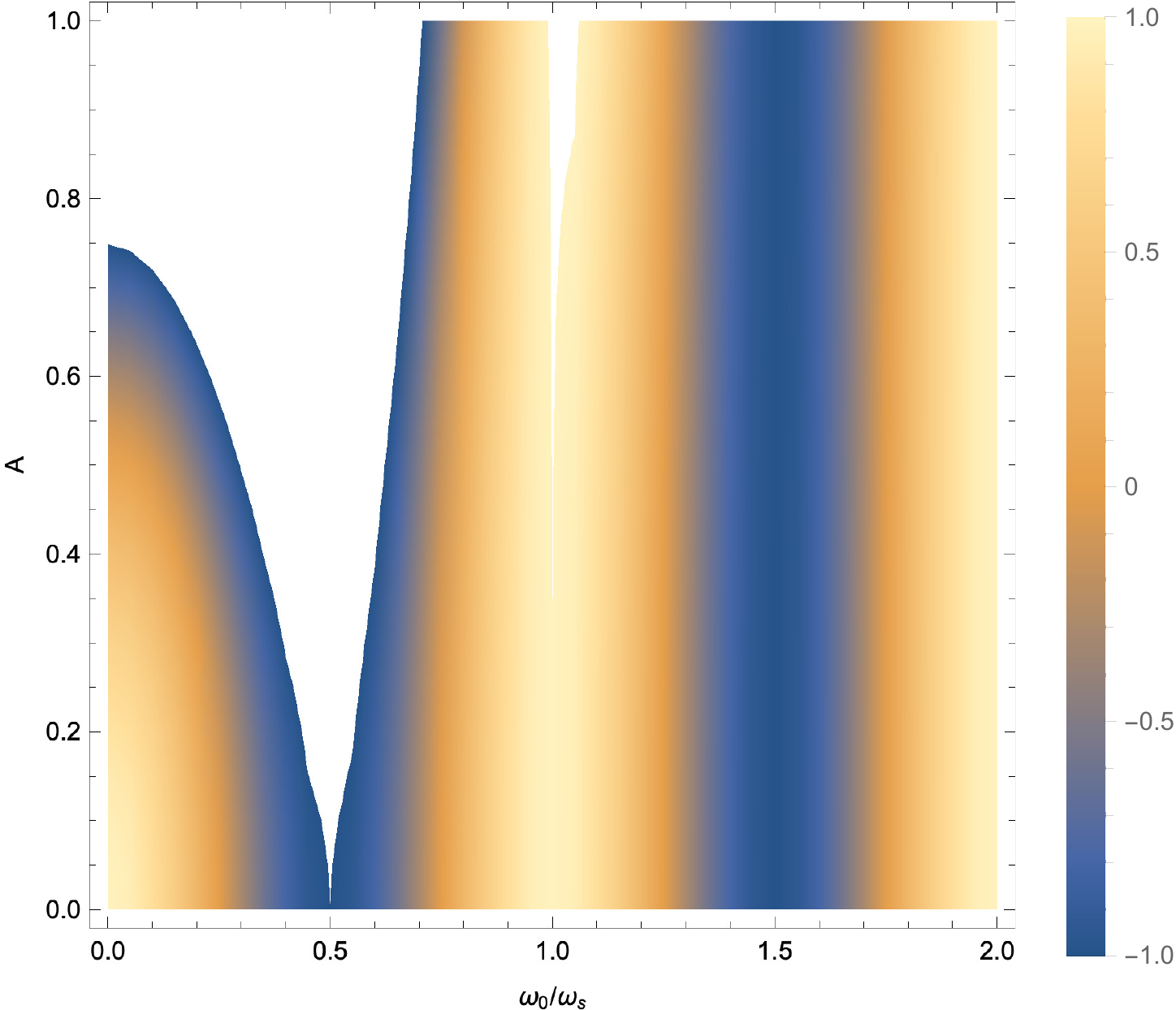}
\caption{Stability/instability regions of (\ref{Eq:Mat}). The white band represents the unstable region. The color palette indicates the trace of the monodromy matrix, $\Delta$. For $\vert \Delta\vert<1$ the solutions of the Mathieu equations are stable. For details see \cite{MathieuEq}.}
\label{fig: Mathieu}
\end{figure}

It is well-known that this equation has unstable solutions in some regions of the parameter space $(a,q)$, which are characterized in terms of the trace of the monodromy matrix of the associated first order system. Note that (\ref{Mathieu}) can be brought to the form (\ref{Eq:Mat}) after the following identifications:
\be
x\rightarrow D,~~ t=\frac{\omega_s}{2}t',~~ a=\frac{4\omega_0}{\omega_s^2},~~\text{and}~~q=-\frac{12 C_2 \hat{A}}{\omega_s^2}\,.
\ee

Using standard results in Floquet analysis, it can be shown that for small $\hat{A}$ the unstable regions satisfy the relation $\omega_0/\omega_s=k/2$, with $k$ a natural number. As $\hat{A}$  increases, the unstable band gets wider, and therefore the number of unstable frequencies grows with $\hat{A}$. On the other hand, as $k$ increases, the unstable region gets narrower. As a consequence, to excite higher unstable frequencies one needs values of the amplitude $\hat{A}$ above the linear level. The regions of instability for Eq. (\ref{Eq:Mat}) can be read from Fig. \ref{fig: Mathieu}. For details in the analysis of Mathieu equations see \cite{MathieuEq}.

\section{Parametric instability due to higher order corrections in the effective action}
\label{R-coupling}

In this appendix we show that a term of the form (\ref{eq:coupling theta ricci}) added to the effective action (\ref{effective-action}) is able to explain the parametric resonance of a zero mode with half the frequency of the homogeneous internal excitation, as observed in our lattice field theory simulations.
 
Supplemented with a coupling between the scalar field $\theta$ representing the amplitude of the shape mode and the Ricci scalar $\mathcal{R}$ associated with the induced metric on the string worldsheet, the effective action reads

\begin{equation}
S=\int d^{2}\xi\sqrt{-\gamma}\left[\left(\alpha+\beta\theta\right)\mathcal{R}-\mu+\frac{1}{2}\gamma^{ab}\partial_{a}\theta\partial_{b}\theta-V\left(\theta\right)\right]\,,
\label{eq:action with coupling}
\end{equation}

where $\alpha$ and $\beta$ are real constants and $V\left(\theta\right)=\frac{1}{2}m_{\theta}^{2}\theta^{2}+\mathcal{O}\left(\theta^{3}\right)$, with $m_{\theta}=\omega_{s}$. Varying this action with respect to the coordinates of the string position, $X^{\mu}\left(\xi^{0},\xi^{1}\right)$, we get the following equations of motion:

\begin{equation}
\partial_{b}\left[\sqrt{-\gamma}\left(M^{ab}-2\beta F^{ab}\right)\partial_{a}X^{\mu}\right]=0\,,
\label{eq:general eom}
\end{equation}

where the symmetric tensors $M^{ab}$ and $F^{ab}$ are given by
\begin{equation}
M^{ab}=2\left(\alpha+\beta\theta\right)G^{ab}+\mu\gamma^{ab}+T^{ab}\,,
\label{eq:tensor M}
\end{equation}
\begin{equation}
F^{ab}=\gamma^{ac}\gamma^{bd}\nabla_{c}\nabla_{d}\theta-\gamma^{ab}\nabla_{c}\left(\gamma^{cd}\nabla_{d}\theta\right)\,.
\label{eq:tensor F}
\end{equation}

On the one hand, $G^{ab}=R^{ab}-\frac{1}{2}\gamma^{ab}\mathcal{R}$ in Eq. (\ref{eq:tensor M}) is the Einstein tensor, which vanishes in $1+1$ dimensions. On the other hand, $T^{ab}$ is the energy-momentum tensor of the scalar field: $T^{ab}=\gamma^{ab}V\left(\theta\right)-\frac{1}{2}\gamma^{ab}\gamma^{cd}\partial_{c}\theta\partial_{d}\theta+\gamma^{ac}\gamma^{bd}\partial_{c}\theta\partial_{d}\theta$.

Assuming that $\theta$ is only a function of time, the expansion of Eq. (\ref{eq:general eom}) for $X^{1}=\psi\left(t,y\right)$ to lowest order in $\theta$ and $\psi$ reads

\begin{equation}
\ddot{\psi}-\left(1+\frac{2\beta}{\mu}\ddot{\theta}\right)\psi=0\,.
\label{eq:linearized eq psi}
\end{equation}

Finally, using the approximation $\theta\left(t\right)=\theta_{0}\cos\left(\omega_{s}t\right)$ and assuming that $\psi\left(t,y\right)=D\left(t\right)\cos\left(\omega_{0}y\right)$, we find a Mathieu equation for the amplitude of the zero mode:

\begin{equation}
\ddot{D}\left(t\right)+\omega_{0}^{2}\left[1-\frac{2\beta\omega_{s}^2}{\mu}\theta_{0}\cos\left(\omega_{s}t\right)\right]D\left(t\right)=0\,.
\label{eq:linearized eq d}
\end{equation}

Therefore, amplification occurs for waves of frequency $\omega_{0}=\omega_{s}/2$.


 \color{black}
\section{Domain wall ring collapse}
\label{appendix-ring}

In our discussion on the effective action describing the internal mode excitation 
of a domain wall string, we argued that one can think of the amplitude of this mode 
as a massive scalar field living on the worldsheet of the soliton. Therefore, it seems 
natural to propose a covariant version of the effective action 
for this mode in a generic string configuration as
\beq
\label{EffA}
S = \int{d^2\xi \sqrt{-\gamma} \left[- \mu + \frac{1}{2} \gamma^{\alpha \beta} \partial_{\alpha} \theta \partial_{\beta} \theta - \frac{1}{2} m_{\theta}^2 \theta^2  \right] }\,,
\eeq
where the first term gives the Nambu-Goto action, while the
other two terms describe the evolution of the internal mode
and its coupling to the Goldstone modes encoded on the
induced metric $\gamma$.

Here we will be interested in studying the motion of a 
circular loop parametrized by the spacetime position
\beq
X^{\mu}(t,\tilde \sigma) = (t, R(t) \cos(\tilde \sigma/R_0), R(t) \sin(\tilde \sigma/R_0))~.
\eeq
Without any excitation, or in other words, with $\theta=0$ everywhere
in the loop, the previous action will yield the equation
\beq
\dot R^2 = 1 - \left( \frac{R}{R_0} \right)^2\,,
\eeq
whose solution for a static initial configuration with $R(t=0) = R_0$
is given by
\beq
R(t) = R_0 \cos\left( \frac{t}{R_0} \right)\,.
\eeq

This represents the evolution of a static ring that collapses
under its own tension in a time $t_c = \pi R_0/2$.

Let us now consider the case of a ring with some extra
energy due to the presence of an excited internal 
mode. In our language, this means that we will start
with an initial condition with $\theta(t=0) = \theta_*$.

The equations of motion in this case become
\bea
\label{eom-excited-ring}
&&\ddot R(t) = - {{(1-\dot R^2)}\over R} \times \left[ {{1}\over{F(\theta)+\frac{\dot{\theta}^{2}}{2\mu}\frac{1+2\dot{R}^{2}}{1-\dot{R}^{2}}}} \left( F(\theta) + {1\over{2\mu}} \dot \theta^2 \left( {{2\dot R^2-1}\over{1-\dot R^2}} \right) + \frac{R \dot R \dot \theta}{\mu} \left( {{\ddot \theta}\over {1- \dot R^2}} + {{m_{\theta}^2 \theta}}\right) \right) \right]\,, \nonumber \\
&&\ddot \theta+\dot \theta {{\dot R}\over {R}} \left(1+ {{R \ddot R}\over {1-\dot R^2}}\right) +  (1-\dot R^2) m_{\theta}^2 \theta =0\,,
\eea
where 
\beq
F(\theta)=\left(1+ \theta^2 {{m_{\theta}^2}\over {2\mu}}\right)\,.
\eeq

These equations are quite complicated, so we will need to solve them
numerically. However, there are a few effects that are worth noticing since
they could be qualitatively important.

We first note that the equation of motion for the radius of the loop
gets corrected by the presence of the scalar field amplitude $\theta(t)$ through
the factor included within large brackets in the equation for $R(t)$ in Eq. (\ref{eom-excited-ring}).
In fact, it can be shown that for time scales larger than the period of oscillation
of the scalar field, this correction decreases the collapsing force on the
ring. This is however difficult to test since this effect becomes relevant 
only for large amplitude configurations for $\theta(t)$, but it is exactly 
in these cases where radiation becomes important and this mechanism
of energy loss is not present in our equations above.

\begin{figure}[h!]
\includegraphics[width=15cm,height=10cm]{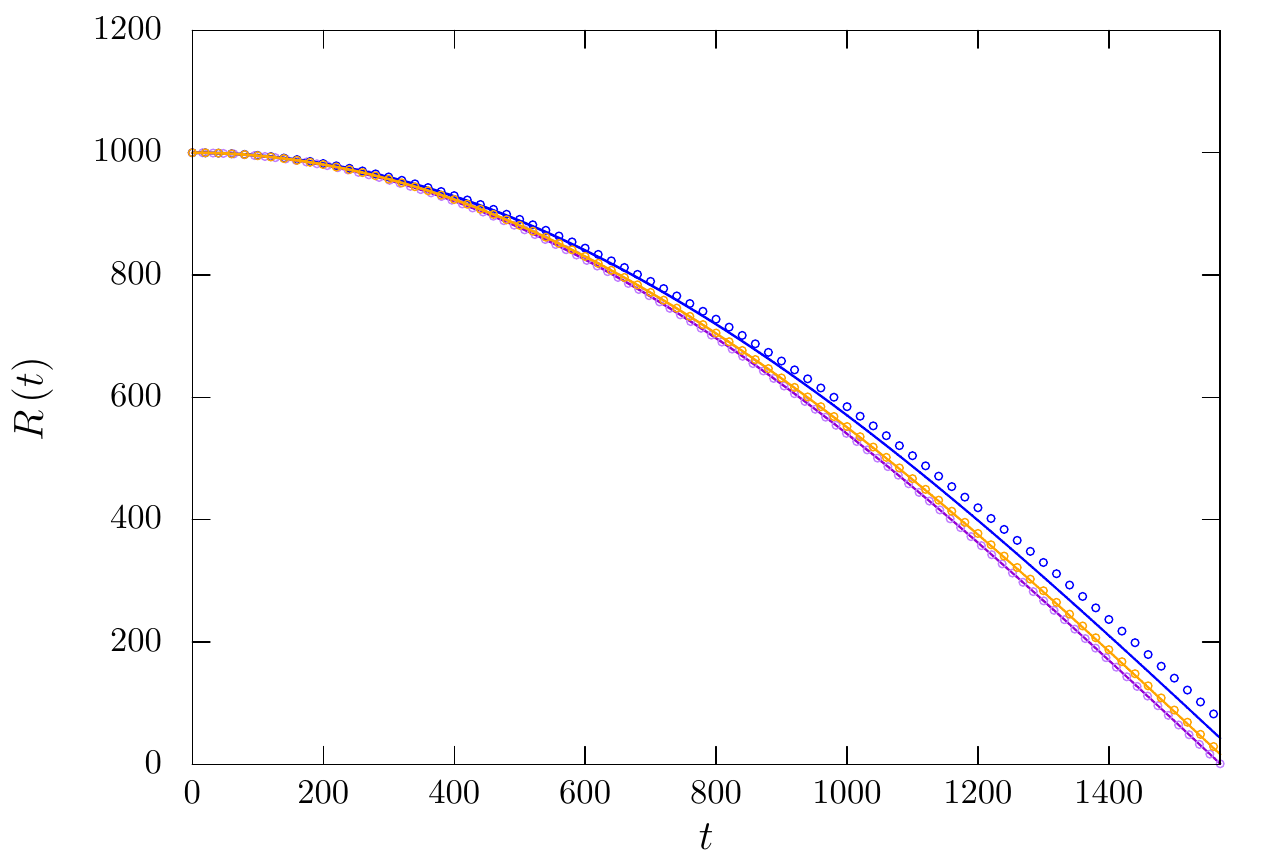}
\caption{Radius of the ring as a function of time for $\theta\left(t=0\right)=0$ (purple),
$\theta\left(t=0\right)=0.2$ (orange) and $\theta\left(t=0\right)=0.4$ (blue). The solid curves are the 
result of the field theory simulation, while the ones with circles correspond to the solution of the 
coupled equations (\ref{eom-excited-ring}).}
\label{fig:ring collapse}
\end{figure}

Another important effect can be seen in the equation for the scalar
field $\theta$. In a collapsing ring, the second term proportional to $\dot \theta \frac{\dot R}{R}$
induces an anti-friction effect that drives the
scalar field to larger amplitudes. This will increase the amount of energy
in the excited mode. However, the collapsing loop also increases its
energy per unit length due to the increase of kinetic energy, so 
the overall effect is not so dramatic as one might think.

In order to test the validity of these equations we have implemented
a field theory simulation that reproduces this collapse from an approximate
initial condition. Here we take advantage of the symmetry of the problem and solve the
equations of motion in the $1+1$ description in terms of the
radial coordinate. We take the initial radius of the domain wall
loop to be very large compared to its thickness. This allows us 
to use the static field configuration for the kink centered at this radius as a pretty 
accurate approximation for the initial state\footnote{We take an absorbing
boundary condition at infinity and a vanishing value of the derivative
of the field at the center of the loop.}.

We show in  Fig. \ref{fig:ring collapse} a comparison of the evolution
of the collapsing ring obtained in field theory with the one predicted
from our effective action described in Eq. (\ref{EffA}). Our results show
a very good agreement between these two descriptions provided that one
starts with a relatively small value of the amplitude of
the shape mode. For larger values one can not neglect the presence
of radiation and the ring trajectory starts to deviate from the effective action
prediction.

\bibliography{dynamics-domain-wall-strings.bib}

\end{document}